\documentclass{ws-ijgmmp}
\begin{document}

\markboth{Pramit Rej and Piyali Bhar}
{Model of hybrid star with baryonic and strange quark matter in Tolman-Kuchowicz spacetime}

\catchline{}{}{}{}{}
%
%%%%%%%%%%%%%%%%%%%%%%%%%%%%%%%%%%%%%%%%%%%%%%%%%%%%%%%%%%%%%%%%%%%%

\title{Model of hybrid star with baryonic and strange quark matter in Tolman-Kuchowicz spacetime}

\author{Pramit Rej \footnote{Corresponding author}}
\address{Department of
Mathematics, Sarat Centenary College, Dhaniakhali, Hooghly, \\West Bengal 712 302, India\\
\email{pramitrej@gmail.com}}

\author{Piyali Bhar}
\address{Department of Mathematics, Government General Degree College, Singur,
Hooghly,\\ West Bengal 712 409, India\\
\email{piyalibhar90@gmail.com , piyalibhar@associates.iucaa.in}}

\maketitle

\begin{history}
\received{(15 December 2021)}
\revised{(17 February 2022)}
\end{history}

\begin{abstract}
The purpose of our present work is to investigate some new features of a static anisotropic relativistic hybrid compact star  composed of strange quark matter (SQM) in the inner core and normal baryonic matter distribution in the crust. Here we apply the simplest form of the phenomenological MIT bag model equation of state $p_q = \frac{1}{3}(\rho_q - 4B_g)$ to correlate the density and pressure of strange quark matter within the stellar interior, whereas radial pressure and matter density due to baryonic matter are connected by the simple linear equation of state $p_r = \alpha \rho - \beta$. In order to obtain the solution of the Einstein field equations, we have used the Tolman-Kuchowicz {\em ansatz} [Tolman, Phys Rev 55:364, 1939; Kuchowicz,  Acta Phys Pol 33:541, 1968] and further derivation of the arbitrary constants from some physical conditions. Here, we examine our proposed model graphically and analytically in detail for physically plausible conditions. In particular, for this investigation, we have reported on the compact object Her $X-1$ [Mass=$(0.98 \pm 0.12)M_{\odot}$; Radius= $8.1_{-0.41}^{+0.41}$ km] in our paper as a strange quark star candidate. In order to check the physical validity and stability of our suggested model, we have performed various physical tests both analytically and graphically, namely, dynamical equilibrium of applied forces, energy conditions, compactness factor, and surface redshift etc. Finally, we have found that our present model meets all the necessary physical requirements for a realistic model and can be studied for strange quark stars (SQS).
\end{abstract}


MSC 2020 Number(s): 85A05, 85A15

%\end{abstract}

%\keywords{f(T) gravity, gravastar, conformal motion}
%\maketitle

%\title{}

%\begin{pacs}
%04.40.Nr, 04.20.Jb, 04.20.Dw
%\end{pacs}

%\footnotetext[0]{\hspace*{-3mm}\raisebox{0.3ex}{$\scriptstyle\copyright$}2013
%Chinese Physical Society and the Institute of High Energy Physics
%of the Chinese Academy of Sciences and the Institute
%of Modern Physics of the Chinese Academy of Sciences and IOP Publishing Ltd}%

%\begin{multicols}{2}

%\maketitle

%\maketitle

\section{Introduction}

Massive neutron stars (NS) are thought to be particularly intriguing astronomical phenomena in a variety of scientific disciplines and study fields, like gravitational physics \cite{Burns2019}, nuclear astrophysics \cite{glendenning2012}, and nuclear particle physics \cite{Heiselberg2000}. This is due to the results of a number of  investigations into their internal properties and the observation of complex occurrences in their properties. With the recent astrophysical observations of heavy neutron stars \cite{demorest2010} and neutron star collisions \cite{TheLIGOScientific2017}, the interest in this study has been substantially increased. These investigations have confirmed the existence of two Neutron stars (NS) having mass about $2M_{\odot}$ \cite{demorest2010,Antoniadis2013,Fonseca2016}.
So many investigations have been performed to demonstrate the possibility of quark matter (QM) in massive NS \cite{Buballa2014,Weber2014,Orsaria2013,Drago2013,Alvarez-Castillo2016,Alford2015}.

{\bf  There are numerous pioneering works on black holes in extended gravity by well-known researchers in astrophysics \cite{nas1,nas2,nas3,nas4,nas5,nas6,nas7,nas8,nas9}(and further references therein).}
 After black holes, NSs represent a unique phenomenon of celestial objects for investigating the densest objects in our Universe (the baryon density $\rho_B \approx 10^{15}\textendash 10^{17} kg/m^3$) \textemdash ~ especially due to the recent and ongoing gravitational wave observations in the LIGO/Virgo events \cite{TheLIGOScientific2017,Abbott2018}. Therefore, they can be considered a superb cosmic laboratory to investigate and take into account the alternative theory of gravity and areas of nonconventional physics. Neutron stars are composed of only nuclear matter. Basically, two types of compact stars are found in our universe, and their cores are made up of quark matter. The first one is called the strange quark stars (SQS), and the second one is known as the hybrid stars. SQS are hypothetical stellar objects that can be considered as ultra-dense NSs. Itoh \cite{Itoh1970} was the first to propose the possible existence of quark stars(QS). Later, Bodmer \cite{Bodmer1971} discovered that strange quark matter (SQM) composed of three types of quarks, namely up (u), down (d), and strange (s) quarks, which is more stable than any ordinary matter. This composition of quark or hybrid neutron-quark stars could be quark matter in part or in whole. {\bf The incorporation of SQM into the fluid distribution plays a crucial role in the formation of ultra dense strange quark objects.} Alford \cite{Alford2001} put forward the idea that adequately high density and low temperature in the
dense core of NS are sufficient to squash the hadrons into quark matter. The transformation of NS into SQS has been reported by Pagliara et al. \cite{Pagliara2013}. Cheng et al. \cite{Cheng1998} reported the structures of realistic strange stars and discussed the dynamical behaviour of strange stars as well as various observational effects in distinguishing strange stars from neutron stars.\par

The de-confinement transition occurs when the matter density of stars reaches very high values near the centre of the star, and the star core includes quark matter. These compact objects are known as hybrid stars, which have a hadronic outer component that surrounds a quark inner component, or mixed hadron quark \cite{Glendenning1991,Glendenning1992,Lawley2004,Pagliara2007,Dexheimer2009,Coelho2010,Lenzi2012,Orsaria2012,Endo2013,Chen2015,Li2015}. There are various ways to describe this hadron-quark transition which have already been introduced in earlier studies: statistical confinement by using a hybrid quark-meson-nucleon (QMN) model \cite{Benic2015,Marczenko2018}, hadron-quark crossover with pressure interpolation \cite{Masuda2012} by employing a smooth interpolation between the $p(\epsilon)$ curves of the two phases, and the extended linear sigma model (eLSM) \cite{Kovacs2020}.

We can calculate the mass of a NS by solving the Tolman-Oppenheimer-Volkoff (TOV) equations with the relevant EoS as input, which incorporates the theoretical information of our theory on dense matter. The hybrid EoS containing both hadronic matter (HM) and quark matter (QM) is usually obtained by combining the EoSs of both HM and QM within their individual models. At the moment, when the microscopic theory of the nucleonic EoS attains a macroscopic level \cite{Baldo1997,Schulze2006,Li2008}, the EoS for QM is still not properly known for zero temperature and high baryonic density appropriate for NS. A lot of work on this has been done by using several models \cite{Fraga2001,Xu2015,Chakrabarty1991,Li2009,Schertler1999,Chu2016,Zacchi2015,Tian2012,Bhar2015}.

Naturally, there is no astrophysical object composed entirely of purely perfect fluid. So, for the formulation of realistic models of super dense stars, it is also important to incorporate the presence of pressure anisotropy. The theoretical investigations of Ruderman \cite{Ruderman1972} proposed that stellar matter may be anisotropic if the matter density for relativistic stellar models attains very high ranges $\rho > 10^{15} gm.cm^{-3}$. According to these observations, in such massive stellar objects, the radial pressure may not be equal to the tangential one. The physical situations involving pressure anisotropy are very diverse. By the term ``anisotropic pressure" refers to the fact that the radial component of the pressure, $p_r(r)$, differs from the corresponding angular components, $p_\theta(r) = p_\phi(r) \equiv p_t(r)$. (The fact that $p_\theta(r) = p_\phi(r)$ is a direct outcome of spherical symmetry.) In fact, spherical symmetry insists that both terms be strictly functions of the radial coordinate `r'. A scalar field with a non-zero spatial gradient is a well known example of a physical system where the pressure is clearly anisotropic. At the level of special relativity, this anisotropic behaviour of a scalar field occurs already, because it is very easy to exhibit that $p_r - p_t = (d\phi/dr)^2$. Boson stars, hypothetical self-gravitating compact stellar objects occurring as the consequence of the coupling of a complex scalar field to gravity, are such systems involving pressure anisotropy naturally \cite{Jetzer1991,Liddle1993,Mielke2000}. Similarly, the energy-momentum tensor of both electromagnetic and Fermionic fields is anisotropic in nature. Isotropy can be viewed as an additional assumption for the behaviour of fields or fluids modelling the stellar interior.

Bowers and Liang \cite{Bowers1974} have extensively studied the effect of local anisotropy in relativistic spheres using equations of state. Furthermore, it is well known that many types of physical processes, as well as a certain range of density, increase local anisotropy \cite{Herrera1997}. The main causes of pressure anisotropy in matter distribution are viscosity, pion condensation \cite{Sawyer1972}, phase transition in super fluid \cite{1980JETP} or by other physical phenomena. In the stellar interior of NS, pions may be condensed. Sawyer and Scalapino \cite{Sawyer1973} have proposed that due to the geometry of the $\pi^{-}$ modes, anisotropic pressure distributions could be considered to describe the pion condensed phase configuration. Mak and Harko \cite{Mak2001} numerically modelled anisotropic celestial objects in general relativity. In the galactic sense, Binney and Tremaine \cite{1987gady} have considered anisotropy for spherical galaxies from a purely Newtonian perspective. Esculpi et al. \cite{2007GR} performed a comparative analysis of the adiabatic stability of the Einstein field equations' solutions with spherical symmetry for a locally anisotropic fluid in general relativity. In this regard, Refs. \cite{Usov2004,Varela2010,Rahaman2010,Kalam2012,Deb2015} (and further references therein) contain some other relevant works on anisotropic compact stellar objects. Following those earlier studies, we find that anisotropy has an impact on the critical mass for stability and the surface redshift ($z_s$).

From the above motivated facts, our goal is to further extend those investigated results, offering a detailed analysis of the changes in the physical properties of relativistic hybrid neutron stars due to the presence of anisotropy. We hope that our present investigation will provide useful information in the analysis of compact stellar objects, as well as allow us to study the behaviour of matter under strong gravitational fields. {\bf The thermodynamic bag model (tdBag) \cite{Farhi1984} and models of the Nambu-Jona-Lasino type (NJL) \cite{Nambu1961,Nambu1961a,Klevansky1992,Buballa2003} are the two most widely used effective quark matter models in astrophysics. Among these two models, the thermodynamic bag model (tdBAG) is widely used in astrophysics. The new vBag model \cite{Klahn2015} was recently introduced as a useful model for astrophysical studies.  } In this paper, we use the MIT bag model for QM. This MIT bag EoS model has been successfully used by several researchers for modelling of SQS \cite{Brilenkov2013,Maharaj2014,Paulucci2014,Panda2015,Isayev2015,Abbas2015,Lugones2017}. {\bf Although the original MIT bag model approximates the impact of quark confinement, it does not explicitly account for the violation of chiral symmetry, which is a key property of Quantum Chromodynamics (QCD). This is the first phenomenological model which successfully modelled the phase transition of the quark gluon plasma (QGP) state to the hadronic phase.} This simplest form of the MIT bag EoS model is very plausible to explore the equilibrium configuration of a celestial compact object composed of u, d and s quarks. The pressure-density relation for QM is given by using the MIT bag EoS as $p_q = \frac{1}{3}(\rho_q - 4B_g)$, where $\rho_q$ and $p_q$ are respectively density and pressure for QM and $B_g$ specifies the bag constant, defined as the difference between the densities of perturbative and non-perturbative Quantum Chromo Dynamical (QCD) vacuum. {\bf Usually, the bag constant lies within the range 58.9-91.5 MeV/fm$^3$ for massless strange quarks\cite{Farhi1984} and the range of $B_g$ for massive quarks is of 56-78 MeV/fm$^3$ \cite{Stergioulas2003}. However, many other researchers have also considered larger values of $B_g$. The range of values is consistent with the CERN experimental data about QGP and is compatible with the RHIC preliminary results\cite{Heinz2000,Blaizot2001,Burgio2002}. In the original MIT bag model, the value of bag constant
was 55 MeV/fm$^3$. In these articles \cite{Burgio2002, Jaikumar2005, Bordbar2012, Kalam2013, Maharaj2014, Lugones2017, Alaverdyan2017, Jasim2018}, the values of $B_g$ are also chosen within the range 60 - 90 MeV/fm$^3$ for strange quark stars. This bag model was further modified by Leonidov et. al \cite{Leonidov1993} by changing the bag constant to a variable dependent on chemical potential $\mu$ and
temperature $T$.} Here the authors have used the specific form of metric potentials introduced by Tolman and Kuchowicz (usually known as the TK {\em ansatz})\cite{Tolman1939,Kuchowicz1968}. Several authors have successfully employed the TK $ansatz$ to develop a realistic and stable model of compact stars, both in the context of General Relativity and of alternative theories of gravity, and it will be explored in detail in the next sections. Recently, we have modelled a charged compact star in $f (R, T)$ gravity by using the TK metric \cite{Rej2021}. In the present work, we have developed a hybrid star model constituted with SQM and normal baryonic matter in Einstein gravity, employing
two types of EoSs; a simple linear relation between the radial pressure and the matter density for barynoic matter and a well-established phenomenological MIT bag model EoS in the well-studied TK {\em ansatz} background for SQM.

The rest of the paper is organized as follows. In Section \ref{sec2}, we discussed the interior spacetime and formulated the basic field equations. The relations between pressure and density for baryonic matter and SQM are also given. In the next section, we will vividly discuss our proposed model. The values of unknown parameters have been derived by matching our interior spacetime smoothly to the exterior Schwarzschild line element in Section \ref{boundary}. The Section \ref{physicalanalysis} is devoted to the physical analysis of the obtained results, for which we have discussed the regularity of metric coefficients, density, pressure, energy conditions, and equilibrium of forces for hybrid star models. Finally, we have summarized our work for the present star model in Section \ref{discussion}.

\section{Basic Field Equations}\label{sec2}
In Schwarzschild coordinates, $x^{\mu}=(t,\,r,\,\theta,\,\phi)$, a spherically symmetry $4D$- spacetime is described by the line element,
\begin{equation}
ds^{2}=e^{\nu}dt^{2}-e^{\lambda}dr^{2}-r^{2}(d\theta^{2}+\sin^{2}\theta d\phi^{2}),
\end{equation}
where $\lambda$ and $\nu$ are the gravitational potentials that depend on $r$ only, since we have considered the static space-time for our present paper.\\
We present a relativistic hybrid star model with strange quark matter and normal baryonic matter in this paper. The concept of the phase transition from quark matter to nuclear matter inspired the hybrid star \cite{Yan:2012mk,Schertler:1997za,Schertler:2000xq}.
The corresponding energy-momentum tensor of the two-fluid model is written as,
  \begin{eqnarray}
  T_0^0=\rho^{\text{eff}}=\rho+\rho_q ,\label{t1}\\
  T_1^1=-p_r^{\text{eff}}=-(p_r+p_q),\\
  \text{and}~~~
  T_2^2=T_3^3=-p_t^{\text{eff}}=-(p_t+p_q).\label{t3}
\end{eqnarray}
Rahaman et al. \cite{Rahaman:2012tu} used two-fluid model to model galactic halo.\\
In the above three Eqs.~(\ref{t1})-(\ref{t3}), $\rho,\,p_r$ and $p_t$ denote the matter density, radial and transverse pressure respectively of the normal baryonic matter, where as $\rho_q$ and $p_q$ denote the respective matter density and pressure due to quark matter.\\
Assuming Gravitational Units $G=1=c$, the Einstein field equations are obtained as,
\begin{eqnarray}
8\pi(\rho+\rho_q)&=&e^{-\lambda}\left[\frac{\lambda'}{r}-\frac{1}{r^{2}} \right]+\frac{1}{r^{2}},   \label{fe1}\\
8\pi (p_r+p_q) &=&e^{-\lambda}\left[\frac{1}{r^{2}}+\frac{\nu'}{r} \right]-\frac{1}{r^{2}}, \label{fe2}\\
8\pi (p_t+p_q) &=& \frac{1}{2}e^{-\lambda}\left[ \frac{1}{2}\nu'^{2}+\nu''-\frac{1}{2}\lambda'\nu'+\frac{1}{r}(\nu'-\lambda')\right].\label{fe3}
\end{eqnarray}
where, $\prime$(prime) represents derivatives with respect to the radial co-ordinate $r$ and the conservation equation of the anisotropic system is given by,
\begin{eqnarray}
 \frac{d}{dr}p_r^{\text{eff}}&=& -\frac{1}{2}\nu'(\rho^{\text{eff}}+p_r^{\text{eff}})+\frac{2}{r}(p_t^{\text{eff}}-p_r^{\text{eff}}).
\end{eqnarray}

%\section{Relation between pressure and density}
To solve the above field equations (\ref{fe1})-(\ref{fe3}), for normal baryonic matter, let us assume a linear equation of state between the radial
pressure $p_r$ and the matter density $\rho$, i.e.,\begin{eqnarray}\label{eos1}
                                                     p_r &=& \alpha \rho-\beta,
                                                   \end{eqnarray}
                                                   where $\alpha$ and $\beta$ both are positive constants and $\alpha$ lies in the range $0<\alpha<1$ with $\alpha\neq 1/3.$
                                                  Let us also assume that, for the quark matter, the pressure-matter density relation is
given by the MIT bag model equation of state \cite{Cheng1998,Witten1984},
\begin{eqnarray}\label{eos2}
  p_q &=& l(\rho_q-4B_g),
\end{eqnarray}
where $B_g$ is the bag constant of units MeV/fm$^3$ \cite{chodos1974new}. For massless strange quarks, the constant $l$ in Eqn. (\ref{eos2}) is equal to $1/3$ and it takes the value $0.28$ for massive
strange quarks, with $m_s$ = 250 MeV  \cite{Stergioulas:2003yp}. According to Witten \cite{Witten1984} such an EoS describes a fluid
composed by up, down and strange quarks only.
Farhi and Jaffe \cite{Farhi1984} investigated the properties of quark matter in equilibrium with weak interactions, containing comparable numbers of up, down, and strange quarks and found that the Witten conjecture is verified for a bag constant between
the values 57 MeV/fm$^3$ and 94 MeV/fm$^3$ for massless and non-interacting quarks.
Malheiro et al. \cite{Malheiro:2003zw} studied strange quark star model in beta equilibrium at high densities. Arba$\tilde{n}$il and Malheiro \cite{Arbanil2016} investigated the influence of the anisotropy in the equilibrium and stability of strange stars through the numerical solution of the hydrostatic equilibrium where they assumed that strange matter inside the quark stars is described by the MIT bag model equation of
state. In this
work, we consider $a=1/3$ and $B_g$ = 60 MeV/fm$^3$ for all plots and we have calculated the numerical values of some physical parameters for both $B_g$ = 60 MeV/fm$^3$ and $B_g$ = 70 MeV/fm$^3$.
\section{TK metric and model of hybrid star}
To solve the system of equations (\ref{fe1})-(\ref{fe3}) along with the EoS described earlier in eqns. (\ref{eos1}) and (\ref{eos2}), we have used the well known Tolman-Kuchowicz ansatz \cite{Tolman1939,Kuchowicz1968} given by,
\begin{eqnarray}\label{eq13}
\nu(r) &=& Br^2+2\ln D,\\ \label{eq14}
\lambda(r)&=& \ln(1 + ar^2 + br^4),
\end{eqnarray}
where $a$, $B$ and $b$ are constant parameters with units km$^{-2}$, km$^{-2}$ and km$^{-4}$ respectively and $D$ is a dimensionless constant. The above metric potentials are well motivated, because they provide a non singular geometry.\\
Several researchers had already employed the above metric potential in the context of general relativity and modified gravity theory. In the presence of the cosmological constant $\Lambda$, which depends on the radial co-ordinate `r', Jasim et al. \cite{Jasim2018} investigated a singularity-free model for spherically symmetric anisotropic strange stars under Einstein's general theory of relativity. Biswas et al. \cite{Biswas:2019doe} created a strange star model using this metric potential along with the MIT bag model equation of state. However, we will now consider the application of this metric potential in the context of modified gravity. Earlier, Bhar et al. \cite{Bhar2019a} used this metric potential to model compact objects in Einstein-Gauss-Bonnet gravity, Javed et al. \cite{Javed2021} used it to model anisotropic spheres in $f (R, G)$ modified gravity, Biswas et al. \cite{sbiswas2020a} obtained an anisotropic strange star with $f (R, T )$ gravity, Majid and Sharif \cite{Majid2020} obtained quark stars in massive Brans-Dicke gravity, Naz and Shamir \cite{Naz2020} discovered the stellar model in $f(G)$ gravity, Farasat Shamir and Fayyaz \cite{FarasatShamir2020} discovered the model of charged compact star in $f (R)$ gravity, Rej et al. \cite{prej2021} investigated the charged compact star in $f(R,\,T)$ gravity, Bhar et al. \cite{Bhar:2020ukr} studied anisotropic compact star in Rastall gravity. It's worth noting that this metric potential successfully produces a compact stellar model that is singularity-free and meets all of the physical requirements.

Now by solving the field equations with the help of the Tolman-Kuchowicz ansatz, the matter density for the normal baryonic matter is obtained as, \begin{eqnarray}
  \rho &=& \frac{1}{8 (1 - 3 \alpha)}\left[\frac{2 \big(a + 2 b r^2 + (2 a - 3 B + 2 b r^2) (1 + a r^2 + b r^4)\big)}{\pi (1 + a r^2 + b r^4)^2}-24 \beta - 32 B_g \right],
\end{eqnarray}
and consequently the radial and transverse pressure are obtained as,
\begin{eqnarray}
  p_r &=& \frac{1}{4 (-1 + 3 \alpha)}\left[\alpha\Big\{16 B_g+\frac{-a - 2 b r^2 + (-2 a + 3 B - 2 b r^2) (1 + a r^2 + b r^4)}{\pi (1 + a r^2 + b r^4)^2}\Big\}+4 \beta\right], \\
  p_t&=&\frac{1}{8 (-1 + 3 \alpha)}\left[\frac{a - 5 a \alpha - 2 B +
 6 \alpha B + C_1 r^2}{\pi(1 + a r^2 + b r^4)^2}-\frac{a (1 + \alpha) - 2 B + C_2 r^2}{\pi(1 + a r^2 + b r^4)}+8 \beta + 32 \alpha B_g\right].
\end{eqnarray}
where $C_1$ and $C_2$ are constants given by,
\[C_1=(2 - 10 \alpha) b + a (-1 + 3 \alpha) B,~~~C_2=b + \alpha b + B^2 - 3 \alpha B^2.\]
The anisotropic factor $\Delta=p_t-p_r$ is given by,
\begin{eqnarray}
  \Delta &=&\frac{ r^2 \big\{a^2 - a B + B^2 + a (2 b + B^2) r^2-
   b \big(1+2 B r^2-(b + B^2) r^4\big)\big\} }{8 \pi (1 + a r^2 + b r^4)^2}.
\end{eqnarray}
The matter density and pressure due to the quark matter
are obtained as,
\begin{eqnarray}
  \rho_q &=& \frac{1}{8 (-1 + 3 \alpha)}\left[\frac{3 \Big\{2 a \alpha +
   4 \alpha b r^2 + \big(a + a \alpha - 2 B + (1 + \alpha) b r^2\big) (1 +
      a r^2 + b r^4)\Big\}}{\pi (1 + a r^2 + b r^4)^2}-24 \beta - 32 B_g\right], \\
  p_q &=& \frac{1}{8 (-1 + 3 \alpha)}\left[\frac{2 \alpha (a + 2 b r^2)}{\pi (1 + a r^2 + b r^4)^2} + \frac{
 a + a \alpha - 2 B + (1 + \alpha) b r^2}{\pi (1 + a r^2 + b r^4)}-8 \beta - 32 \alpha B_g\right].
\end{eqnarray}
\begin{figure}[htbp]
    %\centering
        \includegraphics[scale=.5]{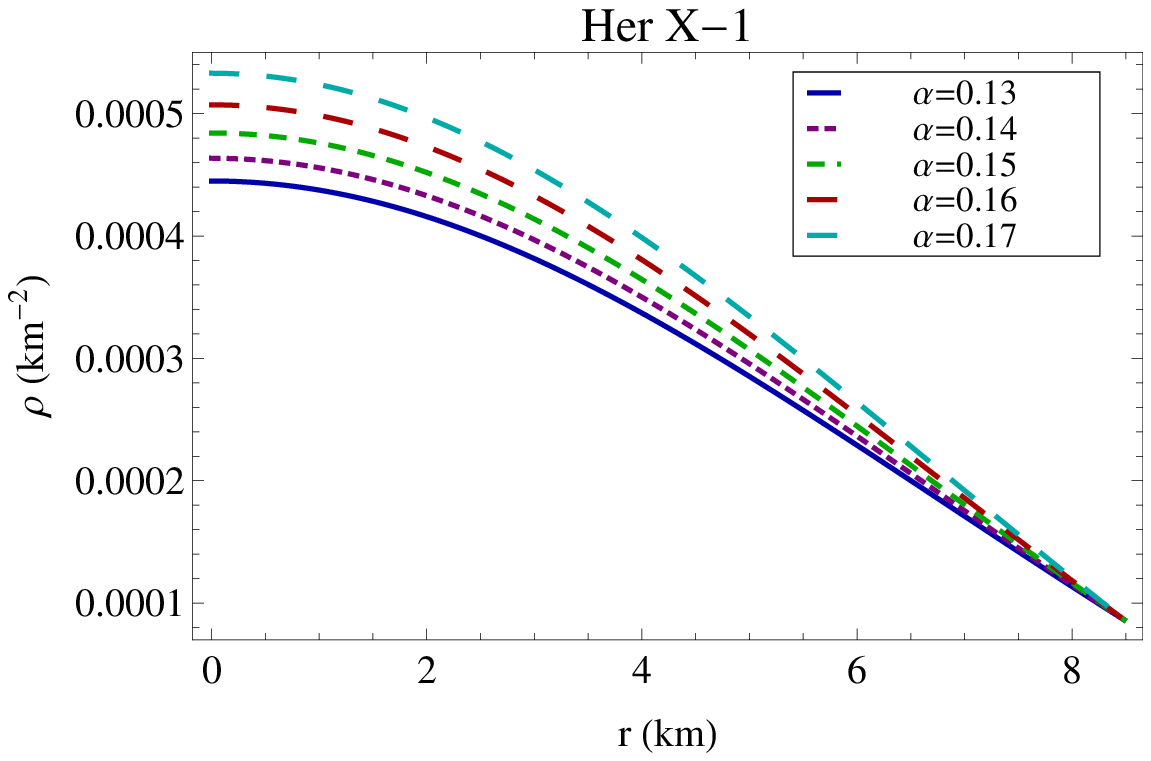}
         \includegraphics[scale=.5]{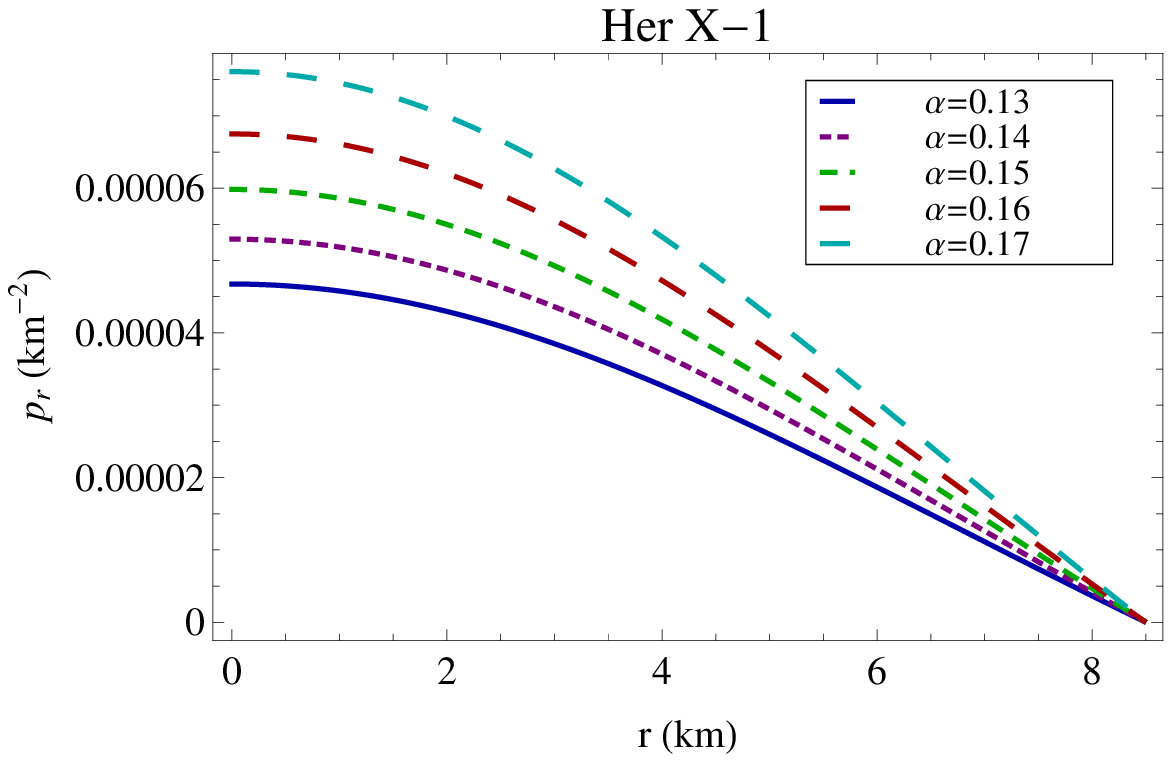}
          \includegraphics[scale=.5]{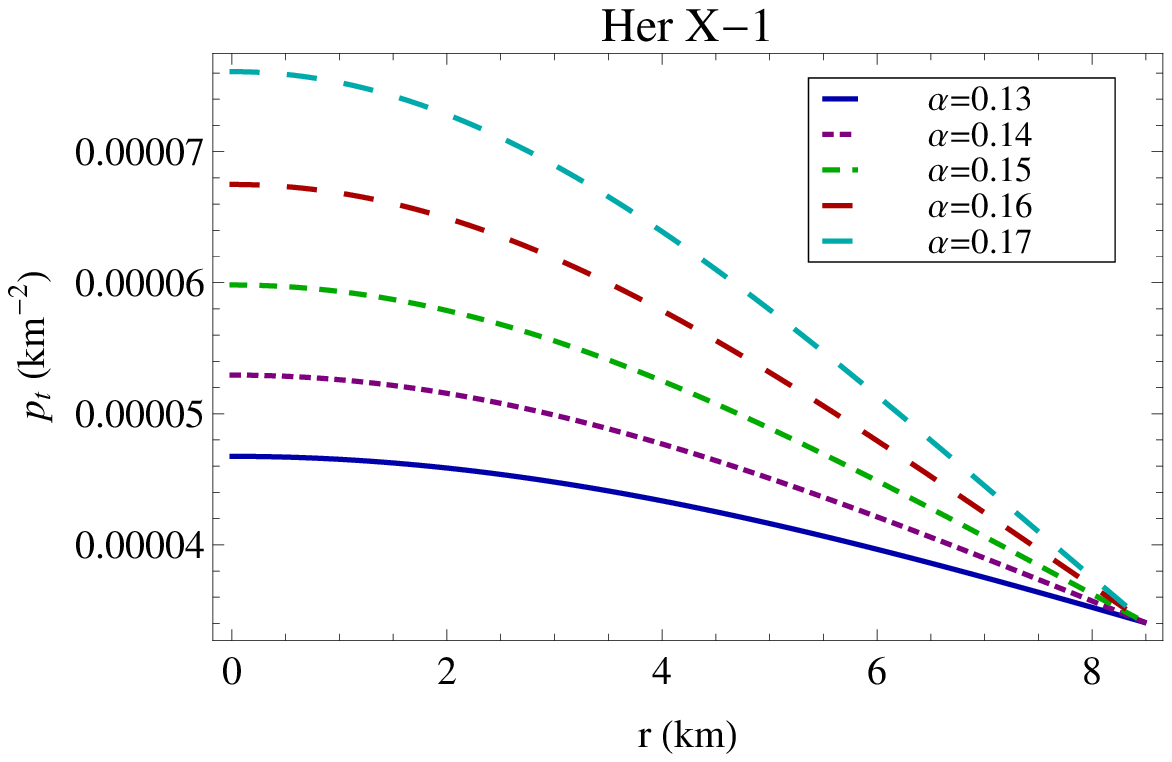}
           \includegraphics[scale=.5]{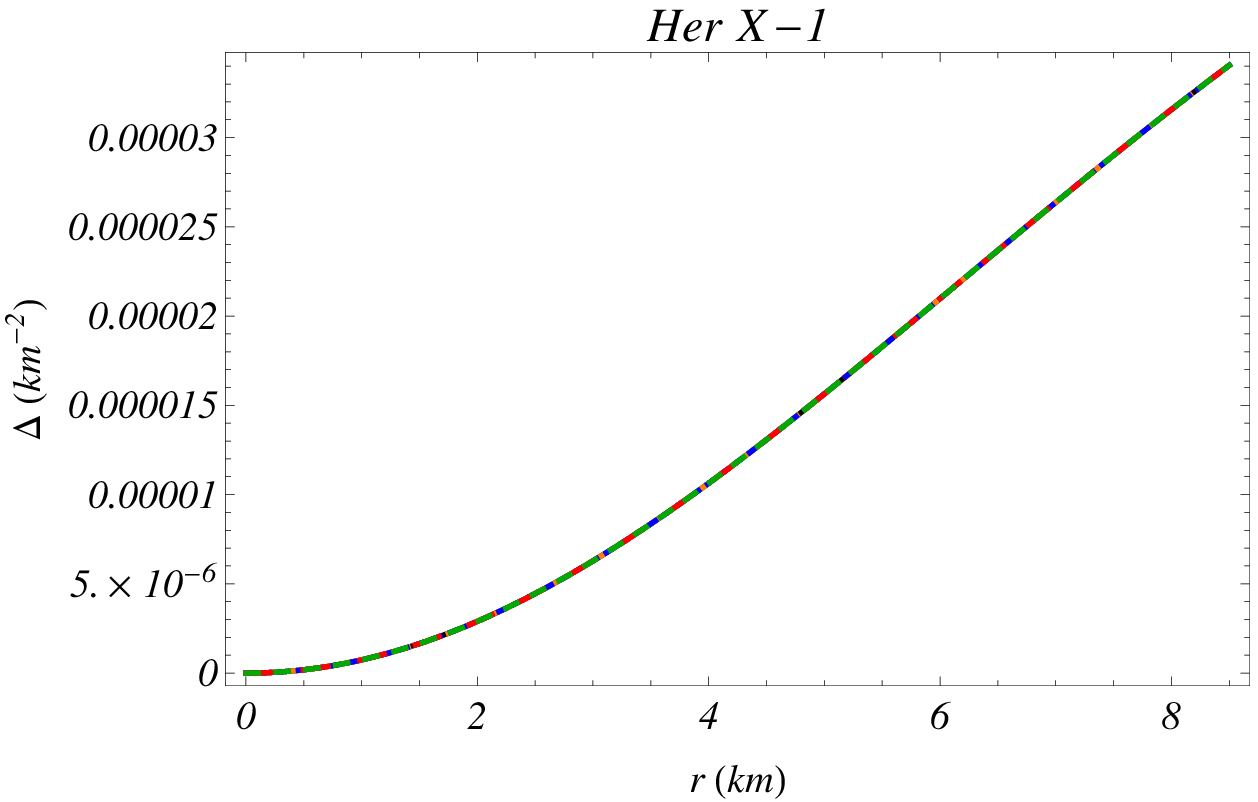}
       \caption{Matter density $\rho$, radial pressure $p_r$, transverse pressure $p_t$ and anisotropic factor $\Delta$ are plotted with varying $r$ inside the stellar interior for Her X-1 for different values of $\alpha$ mentioned in the figures.\label{pr5}}
\end{figure}
 \begin{figure}[htbp]
    %\centering
        \includegraphics[scale=.5]{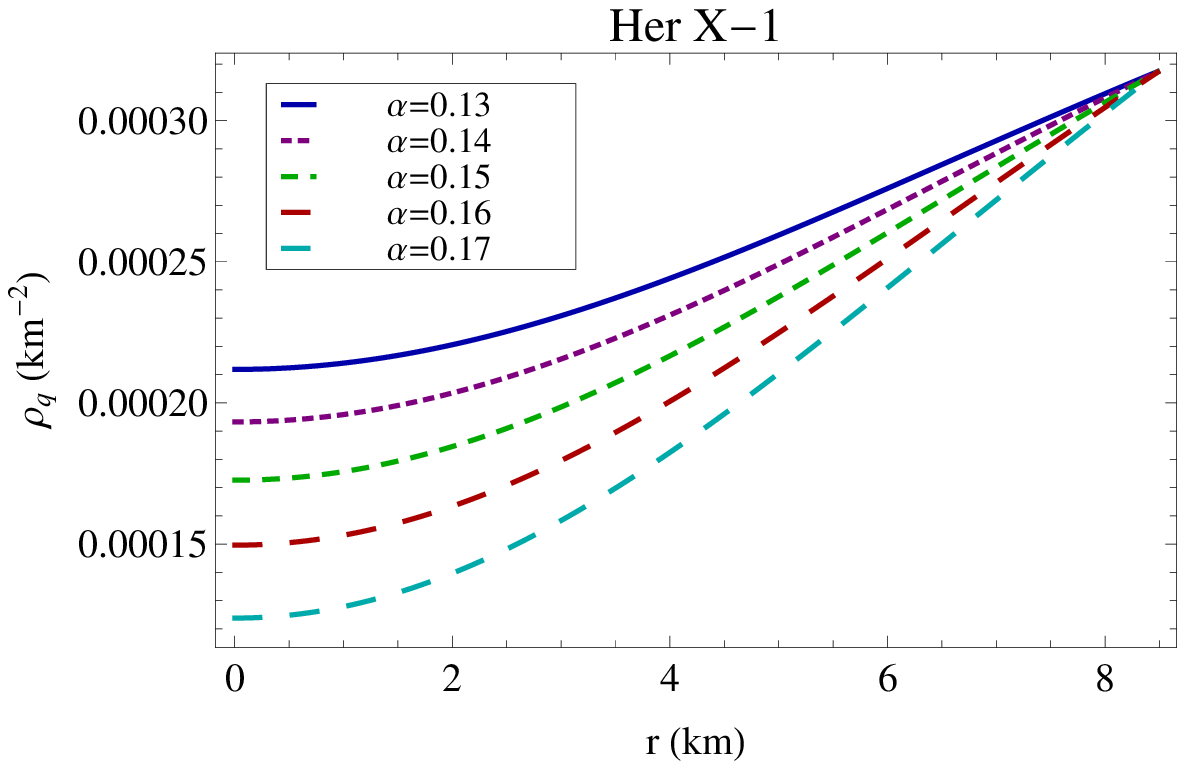}
         \includegraphics[scale=.5]{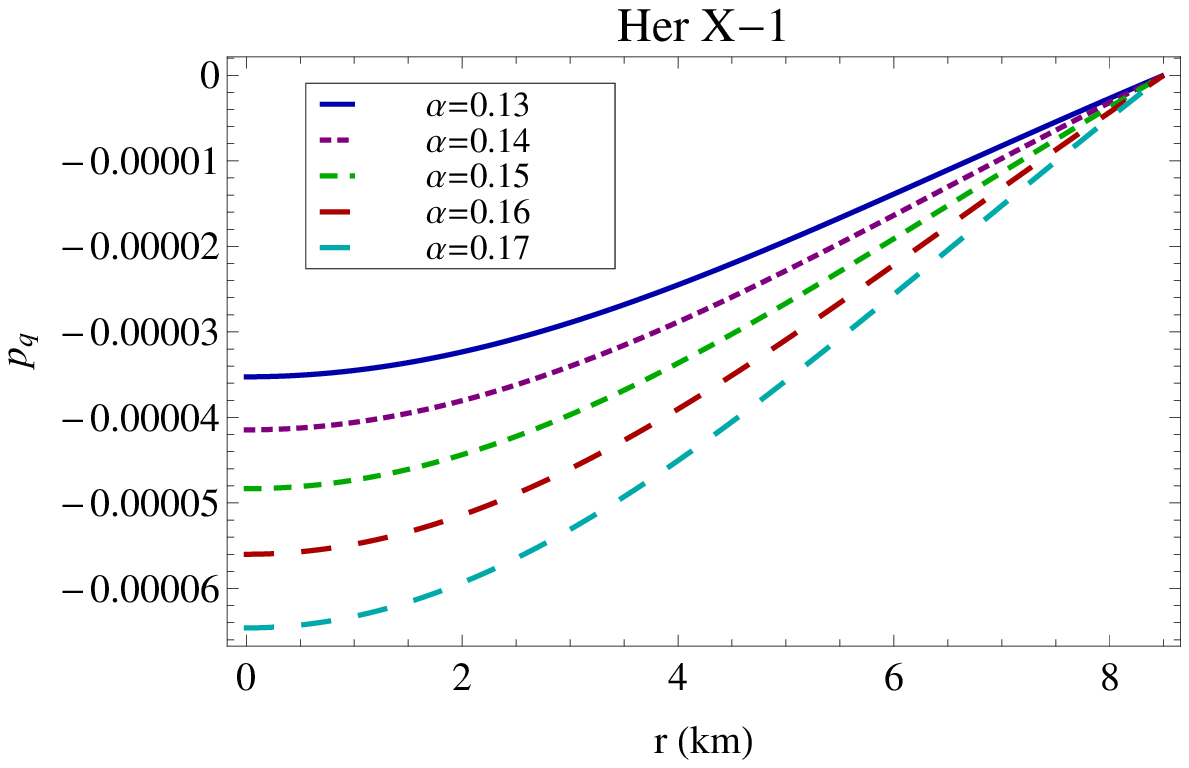}
       \caption{Quark matter density $\rho_q$ and quark matter pressure $p_q$ are plotted with varying $r$ for Her X-1 for different values of $\alpha$ mentioned in the figures.\label{q1}}
\end{figure}

Now the effective density for our present model is obtained as,
\begin{eqnarray}
  \rho^{\text{eff}} =\rho+\rho_q&=& \frac{3 a + (a^2 + 5 b) r^2 + 2 a b r^4 + b^2 r^6}{8\pi(1 + a r^2 + b r^4)^2},
\end{eqnarray}
and the effective radial and transverse pressure are obtained as,
\begin{eqnarray}
  p_r^{\text{eff}}=p_r+p_q&=& \frac{2B-a -b r^2}{8\pi(1 + a r^2 + b r^4)},\\
  p_t^{\text{eff}}=p_t+p_q&=&\frac{-a + 2 B + (-2 b + B (a + B)) r^2 + a B^2 r^4 +
 b B^2 r^6}{8\pi(1 + a r^2 + b r^4)^2}.
\end{eqnarray}
%new
\section{Boundary condition}\label{boundary}
The values of $a,\,b,\,B$ and $D$ must be fixed in order to construct the profiles of the model parameters. To explore the important values of three unknown constants, we match our interior spacetime smoothly to the exterior Schwarzschild line element given by \cite{Schwarzschild1916},
\begin{eqnarray}
ds^{2}&=&f(r) dt^{2}-f(r)^{-1}dr^{2}-r^{2}(d\theta^{2}+\sin^{2}\theta d\phi^{2}),
\end{eqnarray}
along the boundary $r=r_{b}$, where $f(r)=\left(1-\frac{2\mathcal{M}}{r}\right)$ and $\mathcal{M}$ being the mass of the compact star.\\
Now at the boundary $r = r_{b}$ the metric coefficients
$g_{rr}$, $g_{tt}$ all are continuous, which implies,
\begin{eqnarray}
1-\frac{2\mathcal{M}}{r_{b}}&=&e^{Br_b^2}D^2,\label{b1}\\
\left(1-\frac{2\mathcal{M}}{r_b}\right)^{-1}&=&1+ar_b^{2}+b r_b^{4},
\end{eqnarray}
and the continuity of $\frac{\partial}{\partial r}(g_{tt})$ at the boundary gives,
\begin{eqnarray}\label{b3}
\frac{2\mathcal{M}}{r_b^2}=2Br_be^{Br_b^2}D^2.
\end{eqnarray}
Solving the eqns. (\ref{b1})-(\ref{b3}), one can obtain,
\begin{eqnarray}
a&=&\frac{1}{r_b^2}\left[\left(1-\frac{2\mathcal{M}}{r_b}\right)^{-1}-1-b r_b^4\right],\\
B&=&\frac{\mathcal{M}}{r_b^3}\left(1-\frac{2\mathcal{M}}{r_b}\right)^{-1},\\
D&=&e^{-\frac{Br_b^2}{2}}\sqrt{1-\frac{2\mathcal{M}}{r_b}}.
\end{eqnarray}
Also, at the boundary of the star the radial pressure vanishes, i.e., $p_r(r=r_b)=0$, which provides us the expression of $\beta$ as,
\begin{eqnarray}\label{beta}
  \beta&=&-\frac{1}{4}\alpha\Big[16 B_g+\frac{-a - 2 b r_b^2 + (-2 a + 3 B - 2 b r_b^2) (1 + a r_b^2 + b r_b^4)}{\pi (1 + a r_b^2 + b r_b^4)^2}\Big].
\end{eqnarray}
Now from the equation (\ref{beta}), one can note that $\beta$ can not be arbitrarily chosen, it depends on the bag constant $B_g$. So, we have successfully obtained all constants in terms of mass $\mathcal{M}$ and radius ($r_b$) of the strange star.
Using observed values of various candidates of compact stars, we have obtained the numerical values of different constants in the Table \ref{table1}.

\begin{table*}[t]
\centering
\caption{The numerical values of $a,\,B$ and $D$ for some well known compact objects by assuming $b=0.4\times 10^{-5}$~km$^{-4}$.}\label{table1}
\begin{tabular}{@{}ccccccccccccc@{}}
\hline
Star & Observed mass & Observed radius & Estimated  & Estimated &  $a$&$B$&$D$\\
& $M_{\odot}$ & km. & mass ($M_{\odot}$) & radius (km.)& $km^{-2}$ &$km^{-2}$& dimensionless\\
\hline
PSR J1614-2230 \cite{demorest2010} & $1.97 \pm 0.04$ & $9.69 \pm 0.2$   & 1.97& 9.7& 0.0155077& 0.00794205& 0.435751         \\
Vela X-1 \cite{Rawls:2011jw} & $1.77 \pm 0.08$ & $9.56 \pm 0.08$    & 1.77 & 9.5& 0.0131615 & 0.00676124& 0.49463     \\
4U 1538-52 \cite{Rawls:2011jw}& $0.87 \pm 0.07$ & $7.866 \pm 0.21$ & 0.87 & 7.8 & 0.0078171& 0.00403023& 0.72461  \\
LMC X-4 \cite{Rawls:2011jw} & $1.04 \pm 0.09$ & $8.301 \pm 0.2$ & 1.04 & 8.3 & 0.00823644 & 0.004256& 0.685691   \\
Cen X-3 \cite{Rawls:2011jw} & $1.49 \pm 0.08$ & $9.178 \pm 0.13$ & 1.49 & 9.2& 0.0104704 & 0.00540449 & 0.574909 \\
Her X-1 \cite{Abubekerov:2012yj}& $0.85 \pm 0.15$ & $8.1 \pm 0.41$ & 0.85 & 8.5 & 0.00550255 & 0.00289578 & 0.756246  \\
\hline
\end{tabular}
\end{table*}

\section{Physical analysis}\label{physicalanalysis}
In this section, we will look at several physical characteristics of relativistic compact stellar structures. For this purpose we consider the compact star Her X-1 with mass $0.85 M_{\odot}$ and radius $8.5$ km. \cite{Abubekerov:2012yj}. Along with this data we draw the plots of different model parameters and will study different physical features to get more realistic configurations of this stellar structure.

\subsection{Density and pressure profile}
The density and pressure profiles are shown in Fig.~\ref{pr5} for different values of $\alpha$. The figures indicate that they are all monotonic decreasing function of `r', i.e., they have maximum value at the star's centre and the radial pressure $p_r$ vanishes at the boundary. On the other hand, both density and transverse pressure take positive values at $r=r_b$.
The central density $\rho_c$ due to normal baryonic matter of the present model is obtained by,
\begin{eqnarray}
\rho_c=\rho(r=0)=\frac{3(a- B) - 4\pi(3\beta+4B_g)}{4\pi(1-3\alpha)},
 \end{eqnarray}
the central pressure for our present model is obtained as,
 \begin{equation}
 p_c=p_r(r=0)=p_t(r=0)=\frac{3\alpha (a-B) - 4\pi(\beta+4\alpha B_g)}{4\pi(1-3\alpha)}.
 \end{equation}
\begin{figure}[htbp]
    %\centering
        \includegraphics[scale=.4]{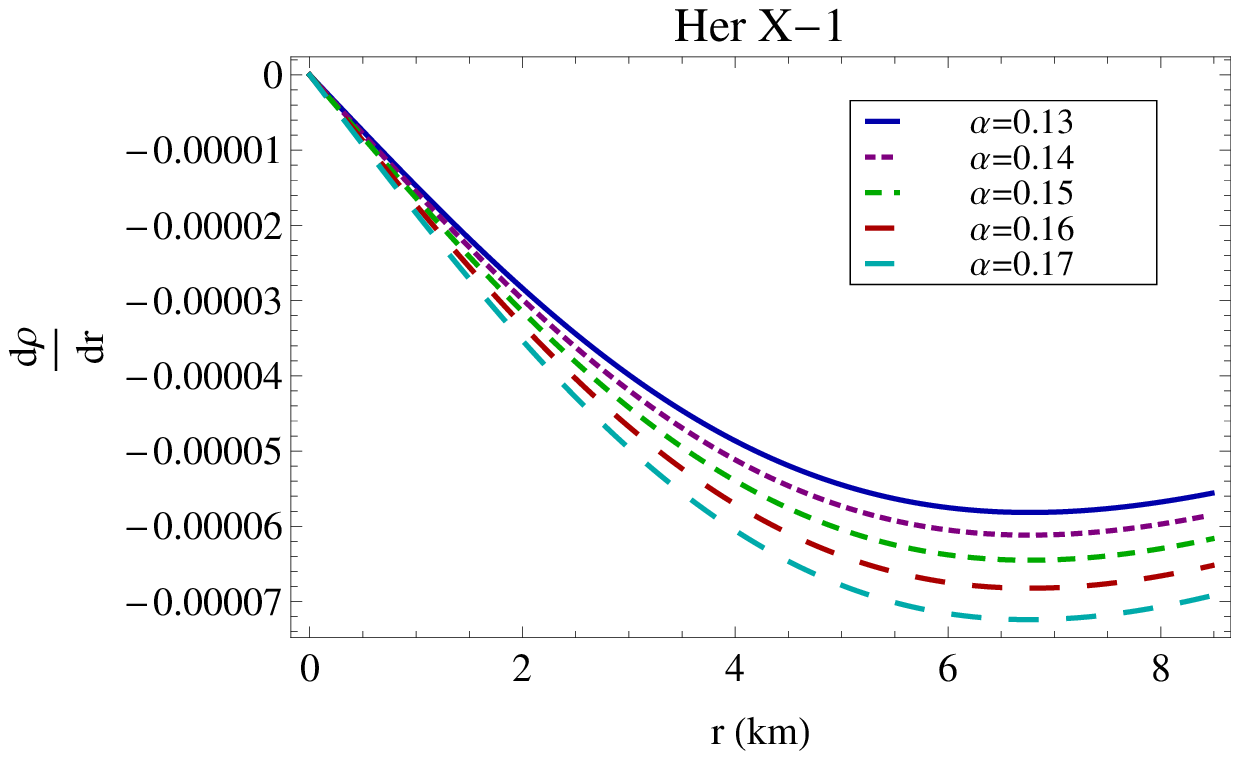}
         \includegraphics[scale=.4]{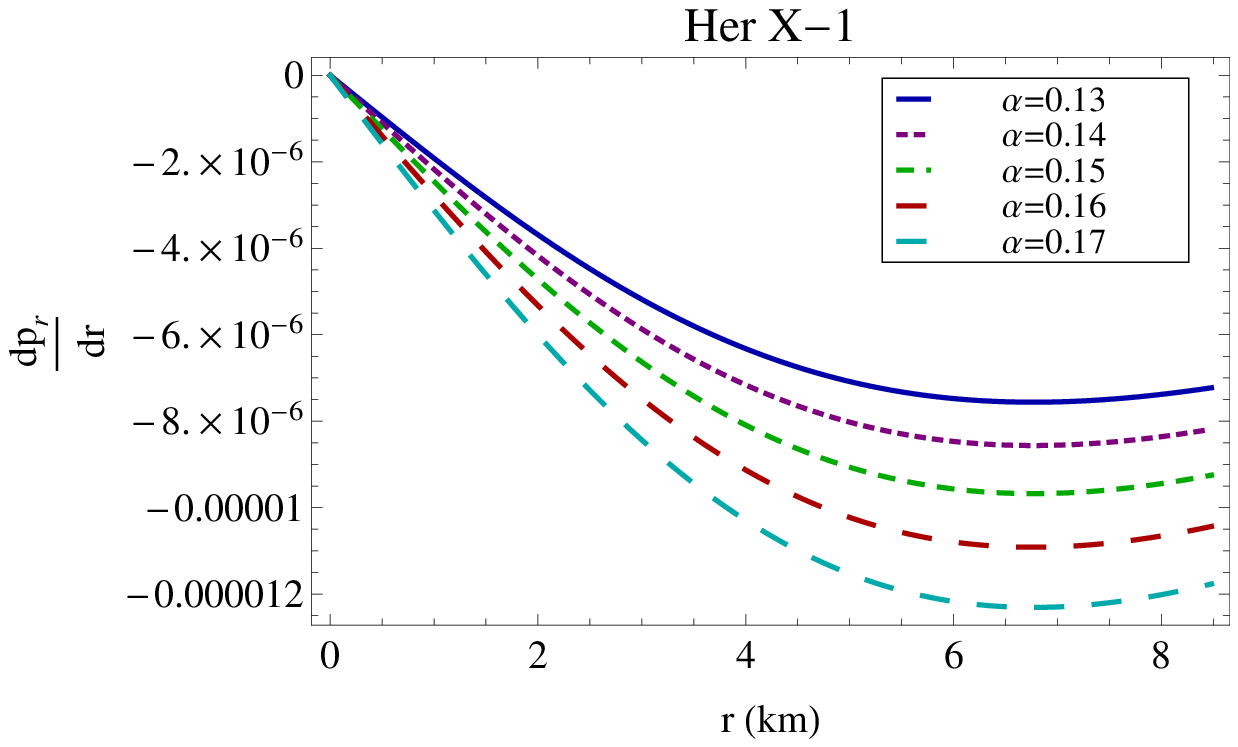}
          \includegraphics[scale=.4]{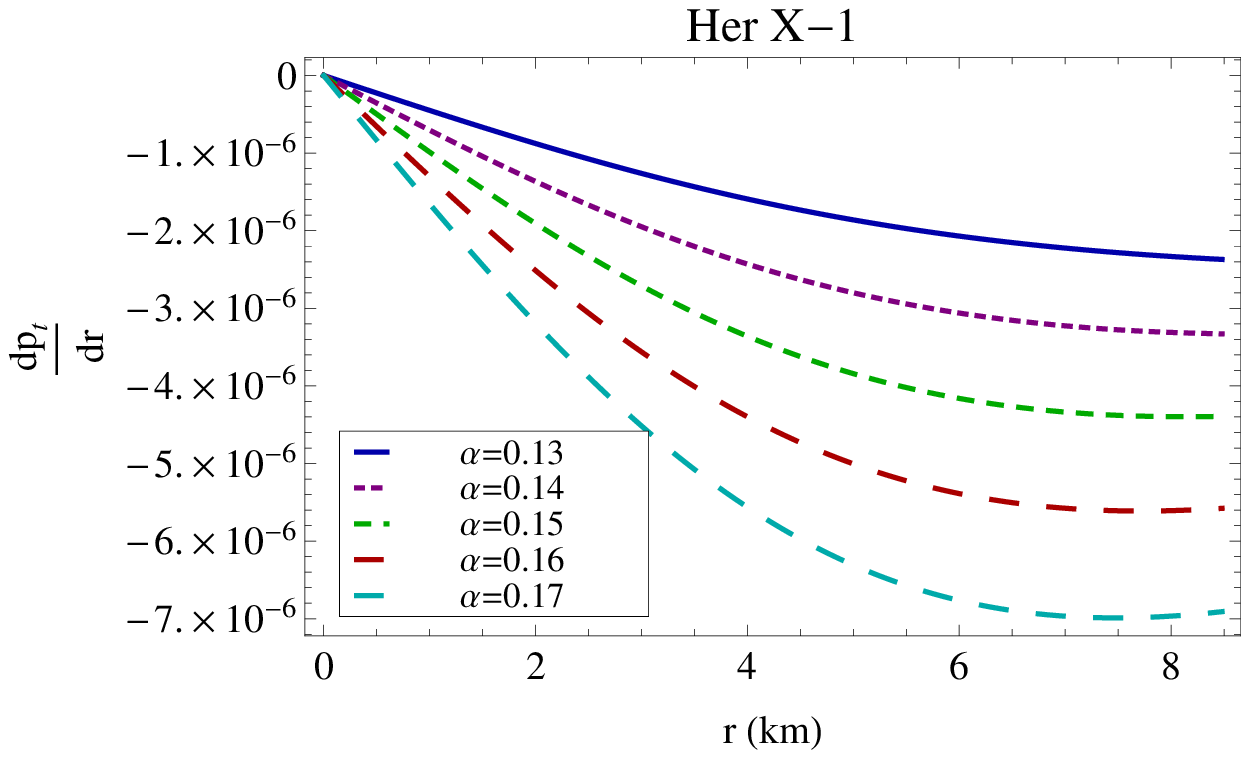}
       \caption{$\frac{d\rho}{dr}$, $\frac{dp_r}{dr}$ and $\frac{dp_t}{dr}$ are plotted against $r$ inside the stellar interior for Her X-1 for different values of $\alpha$ mentioned in the figures.\label{deri}}
\end{figure}
From the figure we also note that, both pressure and density are non-negative inside the stellar interior.
Moreover Zeldovich condition \cite{Zeldovich} indicates that the ratio of pressure to the density is less than $1$,
i.e., $p_r/\rho,p_t/\rho<1$ everywhere within the stellar interior. Now applying the above condition at the center of the star we get, $p_c/\rho_c<1$ which gives us the following relation,
%check
 \begin{eqnarray}
\frac{3\alpha (a-B) - 4\pi(\beta+4\alpha B_g)}{3(a- B) - 4\pi(3\beta+4B_g)}>0
 \end{eqnarray}
It is clear from the Tables.~\ref{table2} and \ref{table3} that the Zeldovich condition for our model is well satisfied.\\
 On the other hand the profiles of pressure and density due to quark matter are shown in Fig.~\ref{q1}. The figure indicates that the quark matter density is monotonic increasing function of radius `r'. The similar nature of quark matter density $\rho_q$ were investigated by Bhar \cite{Bhar2015} in the background of General relativity where as by Abbas and Nazar \cite{Abbas:2021uwt} in the background of $f(R)$ gravity. The above mentioned two works were done by utilizing Krori-Barua {\em ansatz}. Rahaman et al. \cite{Rahaman:2012tu} examined that if $p_q < 0$, with a sufficiently large absolute value, then gravity in the halo is repulsive. For our present model the pressure due to quark matter is also negative.\\
 The density and pressure gradient due to the normal baryonic matter for our present model is obtained as,
\begin{eqnarray}
  \frac{d\rho}{dr} &=&  -\frac{r\big[C_3 + C_4 r^2 +
   3 b C_5 r^4 + 6 b^2 (a - B) r^6 + 2 b^3 r^8\big]}{2 (1-3 \alpha) \pi (1 + a r^2 + b r^4)^3}<0,\\
  \frac{dp_r}{dr} &=&  -\frac{\alpha r\big[C_3 + C_4 r^2 +
   3 b C_5 r^4 + 6 b^2 (a - B) r^6 + 2 b^3 r^8\big]}{2 (1-3 \alpha) \pi (1 + a r^2 + b r^4)^3}<0,\\
  \frac{dp_t}{dr} &=& -\frac{r}{4 (1-3 \alpha) \pi (1 + a r^2 + b r^4)^3}\Big[-2 (a + 2 b r^2) (a - 5 a \alpha - 2 B +
    6 \alpha B + C_6 r^2) + C_6 (1 + a r^2 + b r^4) \nonumber\\&&+ (a +
    2 b r^2) \big(a (1 + \alpha) -
    2 B + C_7 r^2\big) (1 + a r^2 +
    b r^4) - C_7 (1 + a r^2 + b r^4)^2 \Big]<0,
\end{eqnarray}
where, $C_i$'s (i=4,\,5,\,6,\,7) are constants given by,
\begin{eqnarray*}C_3&=&4 a^2 - 4 b - 3 a B,\\
C_4&=&2 a^3 + 8 a b - 3 a^2 B - 6 b B,\\
C_5&=&2 (a^2 + b) - 3 a B,\\
C_6&=&(2 - 10 \alpha) b + a (-1 + 3 \alpha) B,\\
C_7&=&(b + \alpha b + B^2 - 3 \alpha B^2).
\end{eqnarray*}
The profiles shown in Fig.~\ref{deri} investigate the behavior of the variation of radial derivative of density and anisotropic pressures and it is clear from the figure that $\frac{d\rho}{dr},\,\frac{dp_r}{dr}$ and $\frac{dp_t}{dr}$ all took negative value throughout the stellar interior. Moreover at the center of the star $\rho'=0=p_r'=p_t'$, and second derivative of all these variables yields negative results, which confirms about the monotonic decreasing nature of pressures and density.
\subsection{Regularity of the metric coefficients}
\begin{figure}[htbp]
    \centering
        \includegraphics[scale=.45]{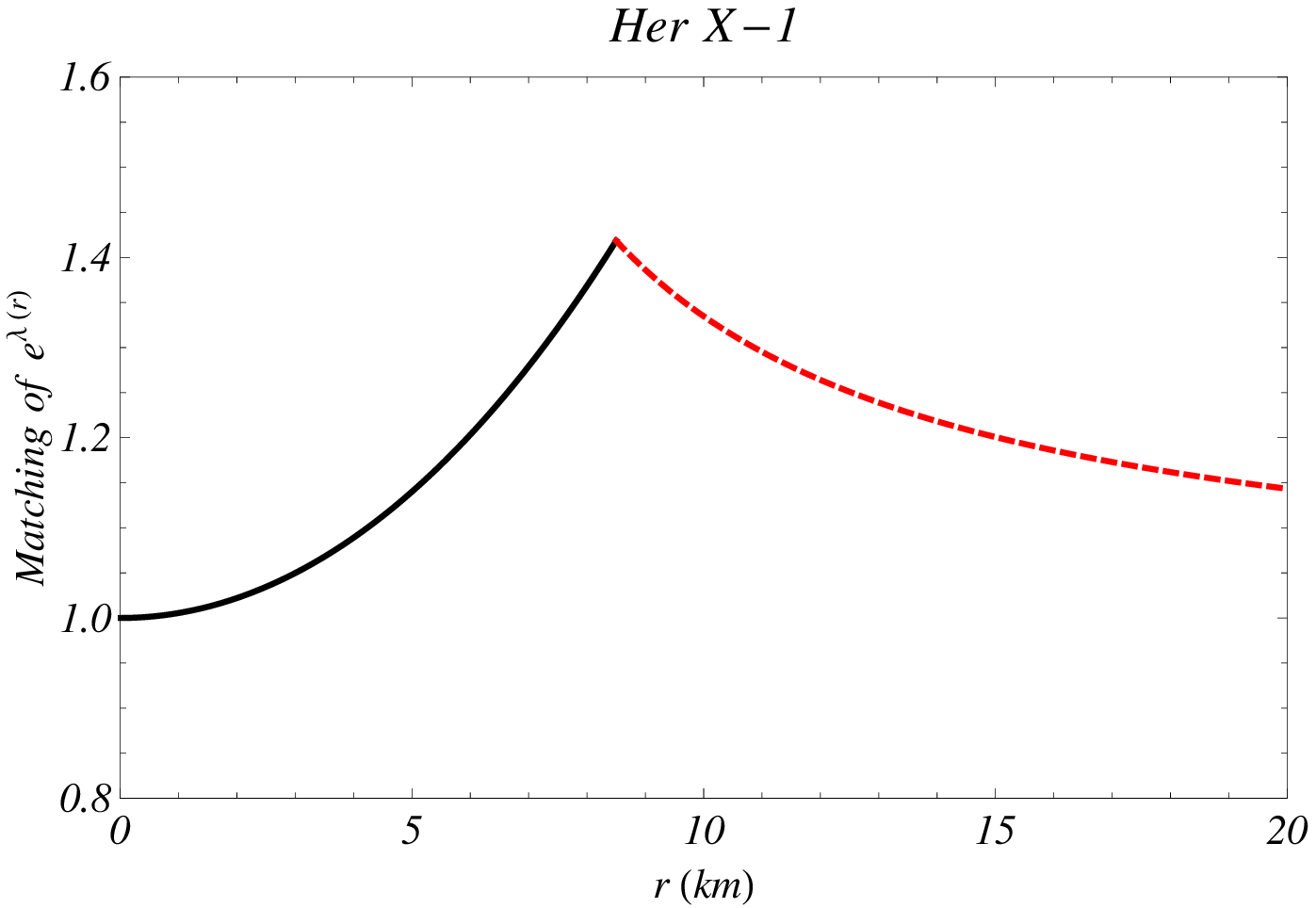}
        \includegraphics[scale=.45]{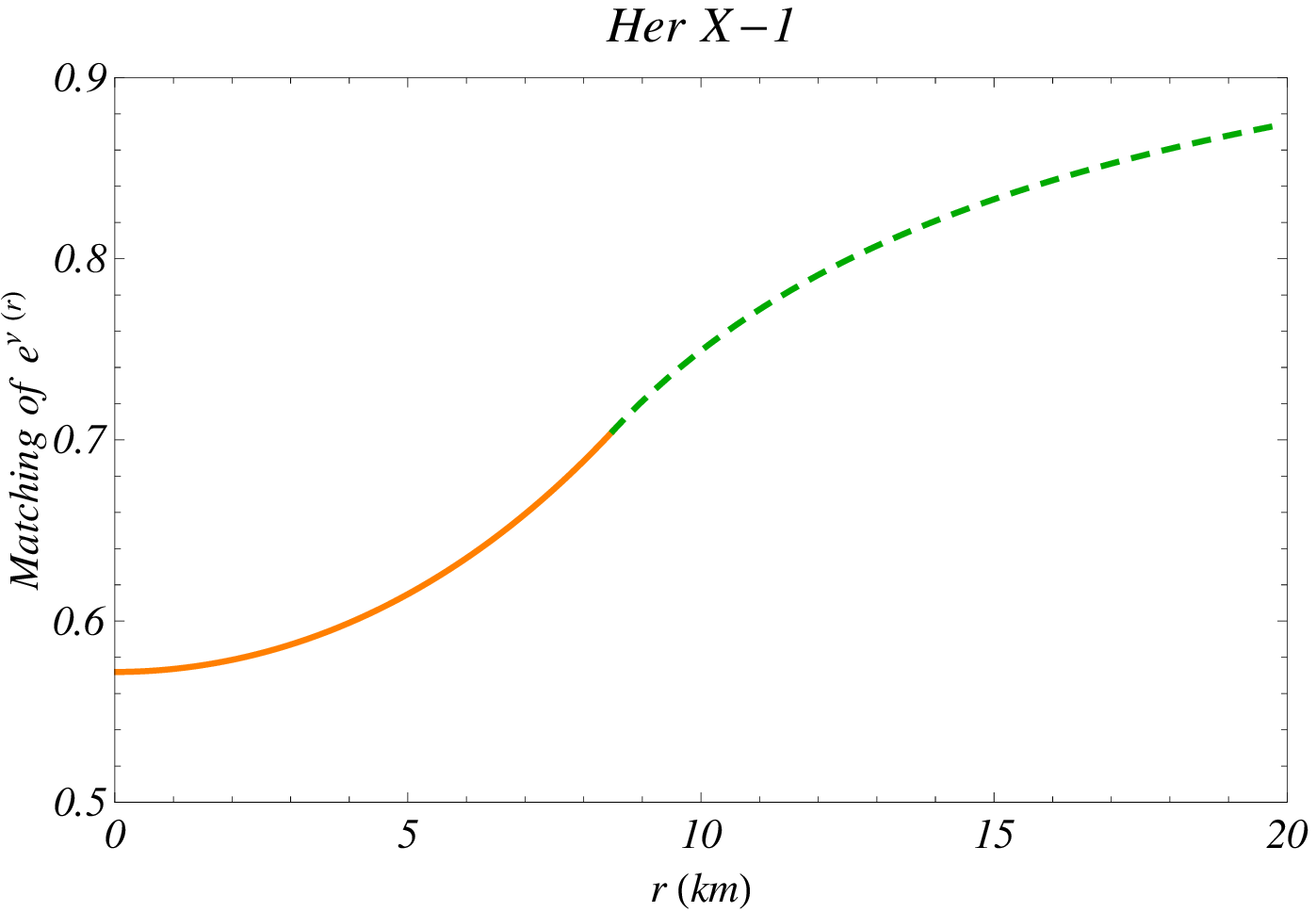}
       \caption{The matching condition of the metric potentials $e^{\lambda}$ and $e^{\nu}$ are
shown against radius for Her X-1. The solid line shows the behavior of the metric potentials in the interior spacetime where as the dashed line shows the nature of the metric coefficients in Schwarzschild geometry. \label{metric}}
\end{figure}

Both the metric potentials are free from singularities inside the radius of the star. Moreover for our present work $e^{\nu(0)}=D^2$, a non-zero constant, and $e^{-\lambda(0)}=1$. The derivative of the metric coefficients give $(e^{\lambda})'=2ar+4br^3$, $(e^{\nu})'=2BD^2re^{Br^2}$. At the center of the star the derivative of the metric potentials are zero. Moreover they are positive and regular inside the interior of the star. The profile of the metric coefficients are shown in Fig.~\ref{metric}. The profiles show a smooth matching of the interior metric potentials to the metric components of the exterior Schwarzschild line element at the boundary.

\begin{table*}[t]
\centering
\caption{The numerical values of $\beta$, central density, surface density, central pressure, compactness ratio, surface redshift and the ratio of central pressure to central density have been shown for different values of $\alpha$ for the compact star Her X-1 by assuming mass $= 0.85~M_{\odot}$, radius $= 8.5 $ km., $b=0.4\times 10^{-5}$~km$^{-4}$ and $B_g$=60 MeV/fm$^3$.}\label{table2}
\begin{tabular}{@{}ccccccccccccc@{}}
\hline
$\alpha$& $\beta$&$\rho_c$&$\rho_s$&$p_c$&$\mathcal{U}$& $z_s(r_b)$& $p_c/\rho_c$\\
\hline
0.13& 0.000011094& $6.0039 \times 10^{14}$& $1.15153\times 10^{14}$ & $5.67727\times 10^{34}$ & 0.0618471 & 0.0682482 &0.105066\\
0.14& 0.0000119478& $ 6.25489\times 10^{14}$ &$1.15153\times 10^{14}$ & $6.43023\times 10^{34}$ & 0.0637102 & 0.0705266 & 0.114226\\
0.15& 0.0000128012&$ 6.53325\times 10^{14}$&  $1.15153\times 10^{14}$ & $7.26532\times 10^{34}$ & 0.0657765&0.0730707&0.123561\\
0.16& 0.0000136546& $6.84373\times 10^{14}$&  $1.15153\times 10^{14}$ & $8.19677\times 10^{34}$ & 0.0680813& 0.0759299&0.133078\\
0.17& 0.000014508&$7.19224\times 10^{14}$&  $1.15153\times 10^{14}$& $9.24228\times 10^{34}$ & 0.0706682 & 0.0791666&0.142782\\
\hline
\end{tabular}
\end{table*}

\begin{table*}[t]
\centering
\caption{The numerical values of $\beta$, central density, surface density, central pressure, compactness ratio, surface redshift and the ratio of central pressure to central density have been shown for different values of $\alpha$ for the compact star Her X-1 by assuming mass $= 0.85~M_{\odot}$, radius $= 8.5 $ km., $b=0.4\times 10^{-5}$~km$^{-4}$ and $B_g$=70 MeV/fm$^{3}$.}\label{table3}
\begin{tabular}{@{}ccccccccccccc@{}}
\hline
$\alpha$& $\beta$&$\rho_c$&$\rho_s$&$p_c$&$\mathcal{U}$& $z_s(r_b)$&$p_c/\rho_c$\\
\hline
0.13& $4.21268 \times 10^{-6}$& $5.28962 \times 10^{14}$& $4.37251\times 10^{14}$ & $5.67727\times 10^{34}$ &0.0458266 & 0.0492383&0.119254\\
0.14& $4.53673 \times 10^{-6}$& $5.54061\times 10^{14}$ &$4.37251\times 10^{14}$ & $6.43023\times 10^{34}$ &0.0476896 & 0.051397&0.128952 \\
0.15& $4.86078 \times 10^{-6}$&$ 5.81897\times 10^{14}$&  $4.37251\times 10^{14}$ & $7.26532\times 10^{34}$ & 0.049756&0.0538069 &0.138729\\
0.16& $5.18483 \times 10^{-6}$& $6.12946\times 10^{14}$&  $4.37251\times 10^{14}$ & $8.19677\times 10^{34}$ & 0.0520607& 0.0565144&0.148586\\
0.17& $5.50888 \times 10^{-6}$&$6.47796\times 10^{14}$&  $4.37251\times 10^{14}$& $9.24228\times 10^{34}$ & 0.0546477 & 0.0595785&0.158525\\
\hline
\end{tabular}
\end{table*}

\subsection{Anisotropic factor}
The anisotropic factor is the difference of radial pressure from the transverse pressure and it is denoted by $\Delta$ and $\frac{2\Delta}{r}$ is termed as the anisotropic force which will be repulsive in nature if $p_t>p_r$ and attractive if the inequalities is in reverse direction. The property of the pressure anisotropy is that it should vanish at the center of the star which indicates that the radial and transverse pressure at the center of the star is equal, in other words at the center, the pressure becomes isotropy. For a physically acceptable model $p_t(r_b)>0$ and $p_r(r=r_b)=0$, therefore $\Delta(r=r_b)$ is always positive. At the same time, it should be positive inside the stellar interior, because positive anisotropic factor creates repulsive force which hold the star against gravitational collapsing as indicated in ref. \cite{Gokhroo:1994fbj}. The nature of pressure anisotropy for our present model is shown in Fig.~\ref{pr5}.

\subsection{Mass radius relation}
\begin{figure}[htbp]
    %\centering
        \includegraphics[scale=.4]{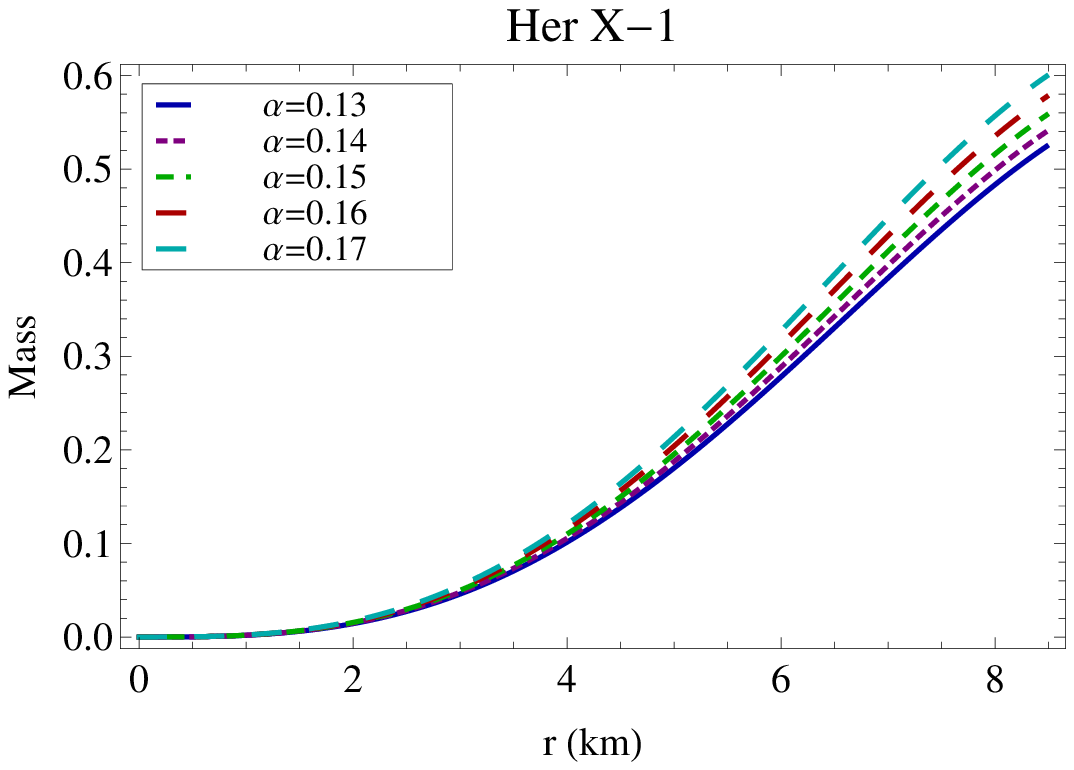}
         \includegraphics[scale=.4]{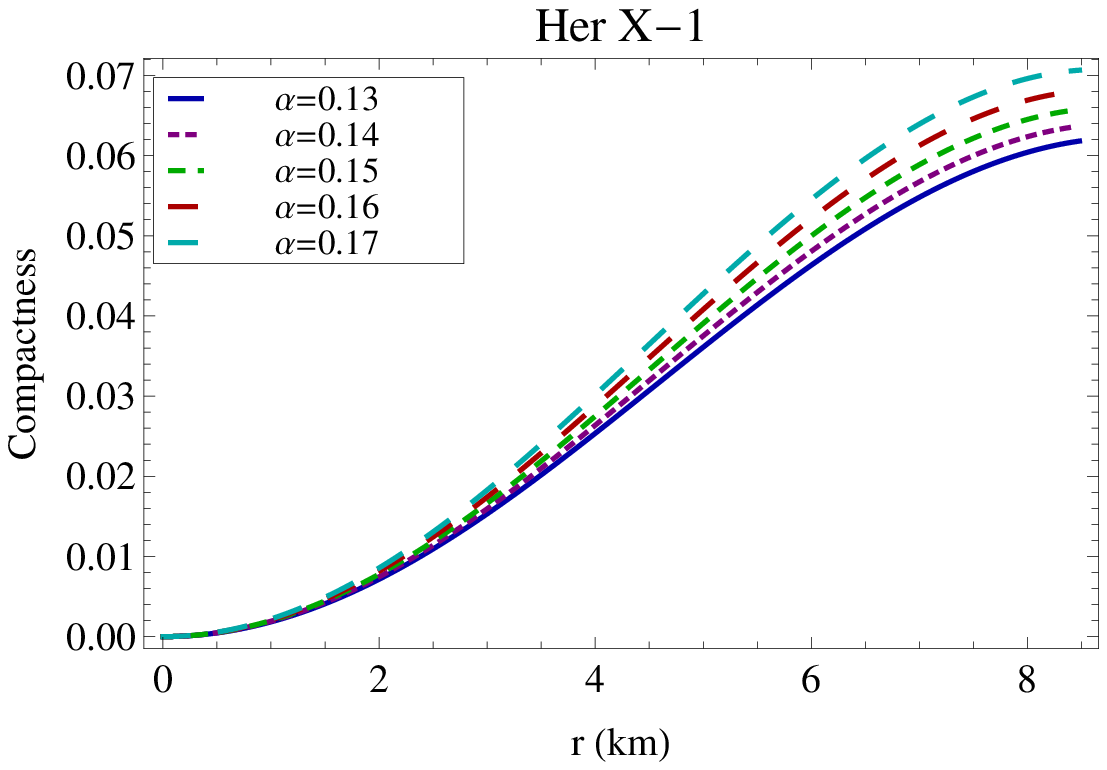}
          \includegraphics[scale=.4]{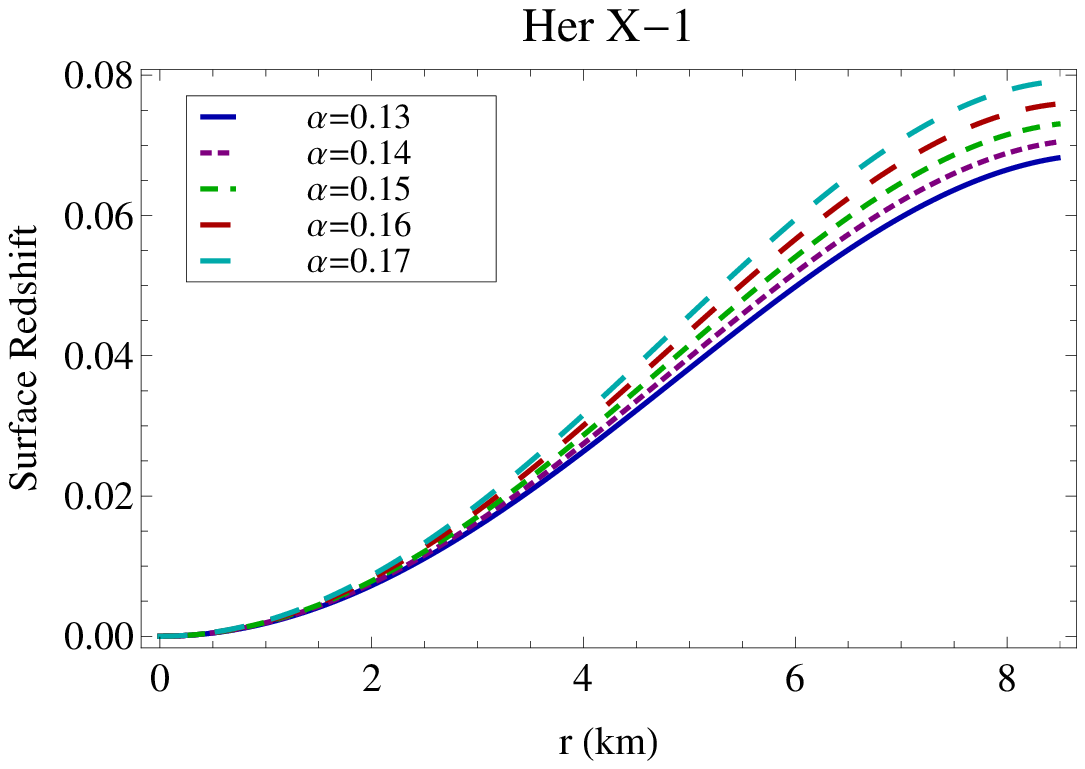}
       \caption{The variation of mass function, compactness and surface redshift are shown against radius $r$ for different values of $\alpha$ mentioned in the figure.\label{mr1}}
\end{figure}

\begin{figure}[htbp]
    %\centering
        \includegraphics[scale=.48]{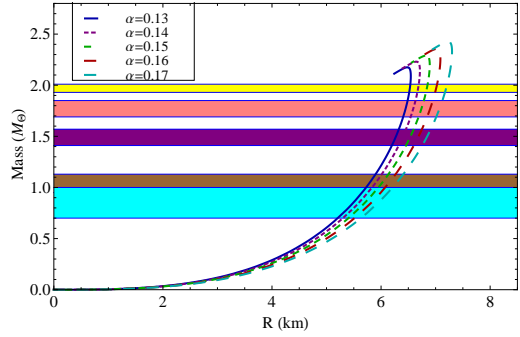}
       \caption{Maximum mass of the system for different values of $\alpha$. The following is a description of the strips: colored strips corresponding to the stars from above (a) Yellow- PSR J1614-2230, (b) Pink- Vela X-1, (c) Purple- Cen X-3, (d) Brown- LMC X-4, and (e) Cyan- Her X-1. \label{mr2}}
\end{figure}

In this subsection, we will look at our model's mass function, which is calculated as follows:
 \begin{eqnarray}
m(r)&=&4\pi\int_0^r \rho(\xi) \xi^2 d\xi,\nonumber\\
&=&\frac{1}{2(1-3\alpha)}\left[\frac{3\sqrt{2}}{E\sqrt{b}}\left\{\frac{(b - (a + E) B)\tan^{-1}\sqrt{\frac{2b}{a+E}}r}{\sqrt{a+E}}-\frac{(b + (-a + E) B)\tan^{-1}\sqrt{\frac{2b}{a-E}}r}{\sqrt{a-E}}\right\}\right.\nonumber\\&&+\left.4r-\frac{8\pi}{3} (3 \beta + 4 B_g) r^3-\frac{r}{1+ar^2+br^4}\right],
\end{eqnarray}
where $E=\sqrt{a^2 - 4 b}$ is a constant. One can note that the mass function depends on the bag constant $B_g$.\\The compactness factor and surface redshift for our present model are respectively obtained as, $u(r)=\frac{m(r)}{r}$ and $z_s=\left(1-2\frac{m(r)}{r}\right)^{-\frac{1}{2}}-1$. The mass function, compactness and surface redshift for our present model are shown in Fig.~\ref{mr1} for different values of $\alpha$. The figures indicate that all the functions are monotonic increasing functions of `r', i.e., they attain maximum values at the boundary of the star. We denote the maximum values of compactness factor and surface redshift as $\mathcal{U}$ and $z_s(R)$ respectively. Here $\mathcal{U}=\frac{M}{r_b}$ and $z_s(r_b)=(1-2\mathcal{U})^{-\frac{1}{2}}-1$ with $M=m(r_b)$. The numerical values of $\mathcal{U}$ and $z_s(r_b)$ for the compact star Her X-1 for different values of $\alpha$ are calculated in the Tables~\ref{table2} and \ref{table3} for two different values of the bag constants.\\
In Fig. \ref{mr2}, we show how the total mass, M (normalised in solar mass $M_{\odot}$), varies with respect to the total radius, R, due to different values of the parameter $\alpha$ mentioned in the figure, where the bag constant is $B_g=$60 MeV/fm$^3$. As the value of $\alpha$ grows, we find that the system's maximum mass increases, as illustrated in the figure.

\subsection{Energy conditions and Equation of state}
 In the field of general relativity, energy conditions play a vital role for a model of compact star. To be physically acceptable, the pressures and density should obey some bound. There are mainly four types of energy conditions namely : dominant energy condition (DEC), strong energy condition (SEC), weak energy condition (WEC) and null energy condition (NEC), and defined as :
\begin{itemize}
\item NEC:~$\rho+p_r \geq 0,~\rho + p_t  \geq 0,$
\item WEC:~$\rho+p_r \geq 0,~\rho + p_t  \geq 0,~ \rho \geq 0,$
\item SEC:~$\rho+p_r \geq 0,~\rho + p_t \geq 0, \rho+ p_r +2 p_t \geq 0,$
\item DEC:~$\rho-p_r \geq 0,~\rho - p_t \geq 0,~ \rho \geq 0$.
\end{itemize}

\begin{figure}[htbp]
    %\centering
        \includegraphics[scale=.4]{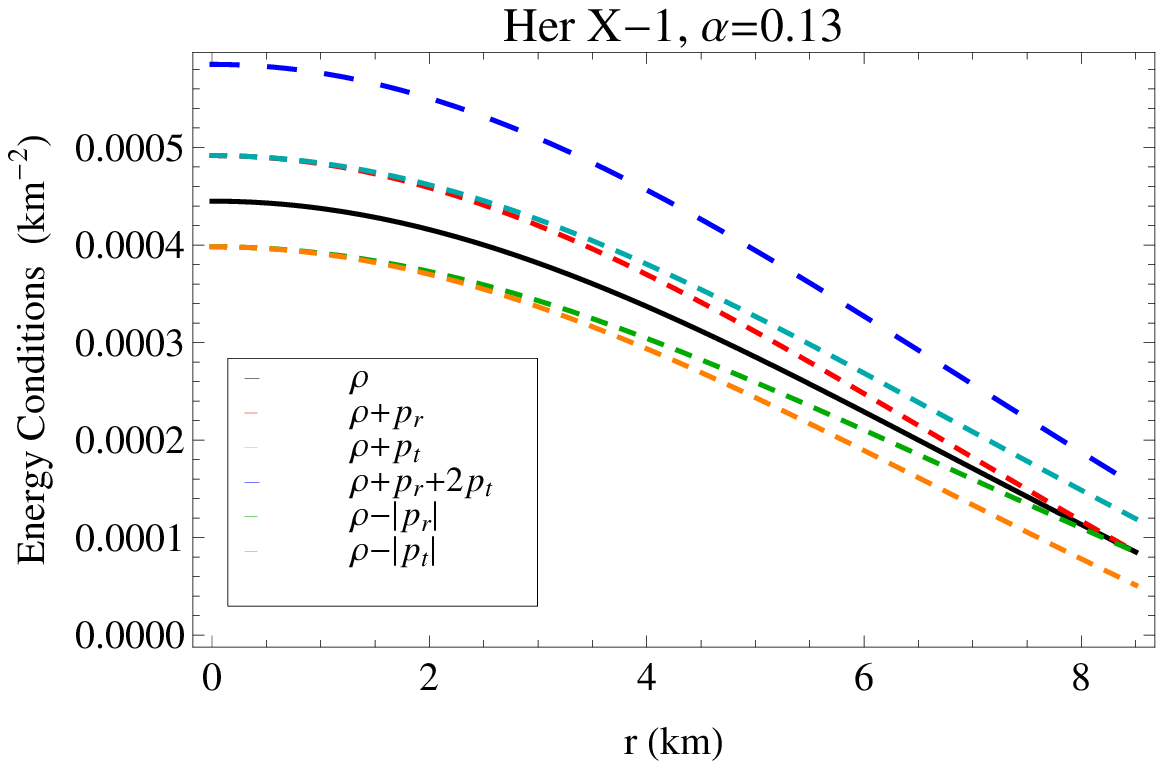}
         \includegraphics[scale=.4]{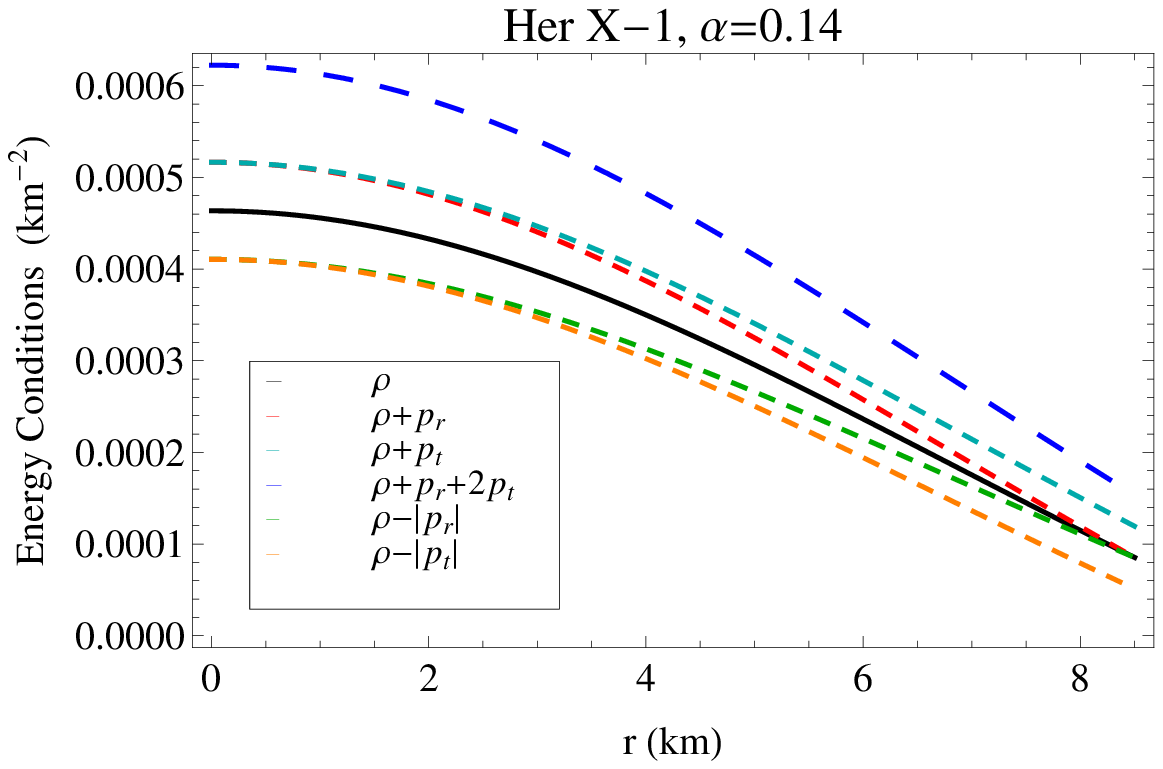}
          \includegraphics[scale=.4]{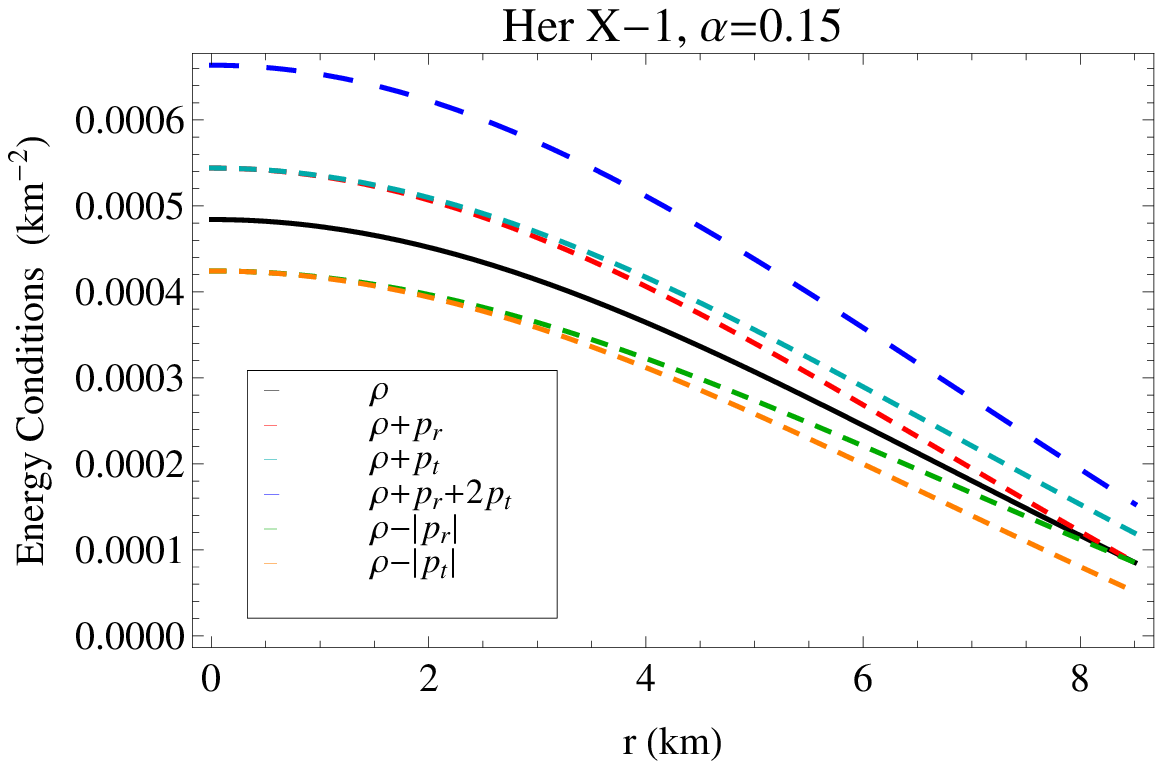}
           \includegraphics[scale=.4]{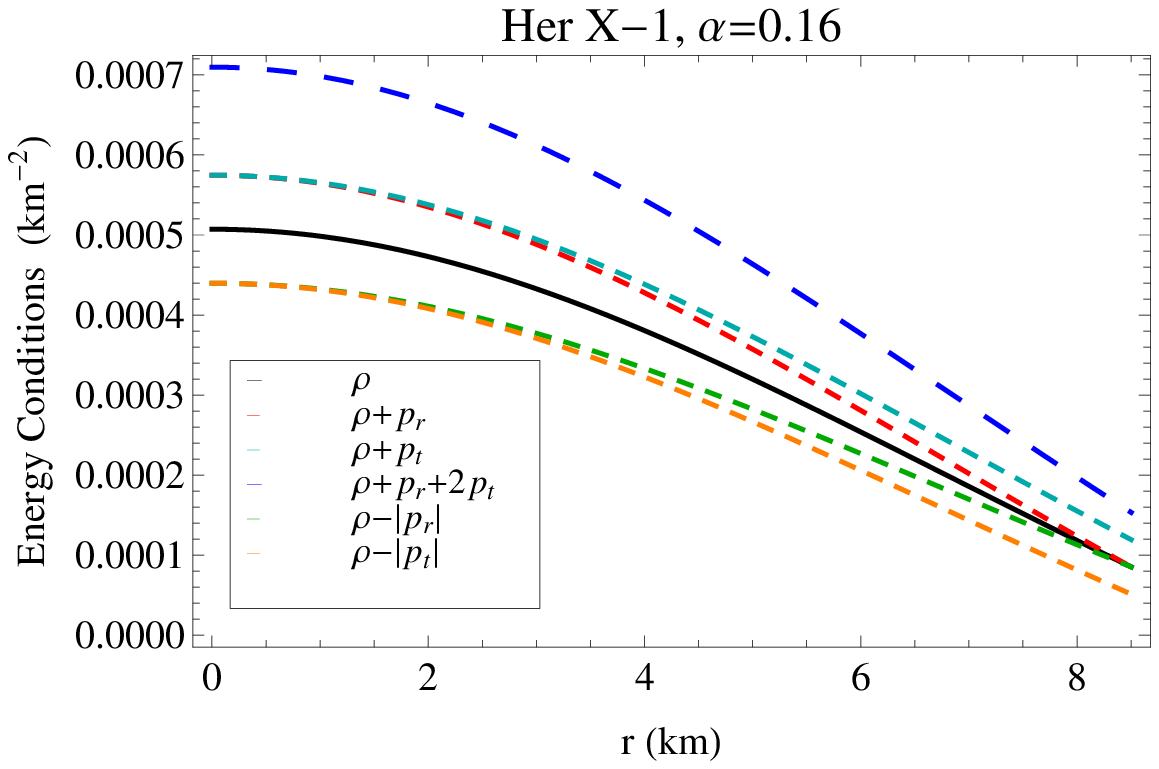}
            \includegraphics[scale=.4]{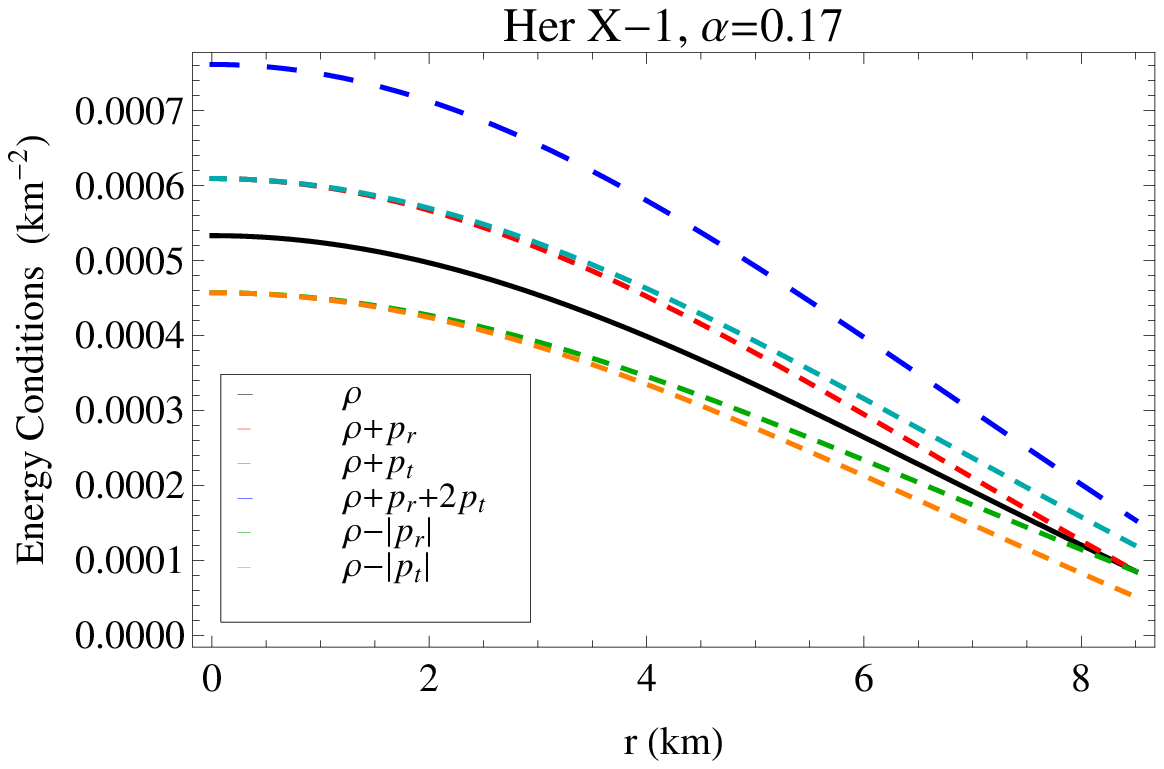}
       \caption{All the energy conditions are plotted inside the stellar interior for Her X-1 for different values of $\alpha$ mentioned in the figures.\label{ec1}}
\end{figure}
With the help of a graphical representation in Fig.~\ref{ec1}, we demonstrated that our present model satisfies all of the energy criteria for different values of $\alpha$. \\

The equation of state parameters $\omega_r$ and $\omega_t$ for a model of  compact star are defined by $\omega_r=\frac{p_r}{\rho}$ and $\omega_t=\frac{p_t}{\rho}$. To check the behavior of equation of state parameter, we have drawn their profiles in Fig.~\ref{eos1}. The figures show that $\omega_r$ is monotonic decreasing function of `r' but $\omega_t$ is monotonic increasing. Both of them lie in the range $0<\omega_r,\,\omega_t<1$.
\begin{figure}[htbp]
    %\centering
        \includegraphics[scale=.5]{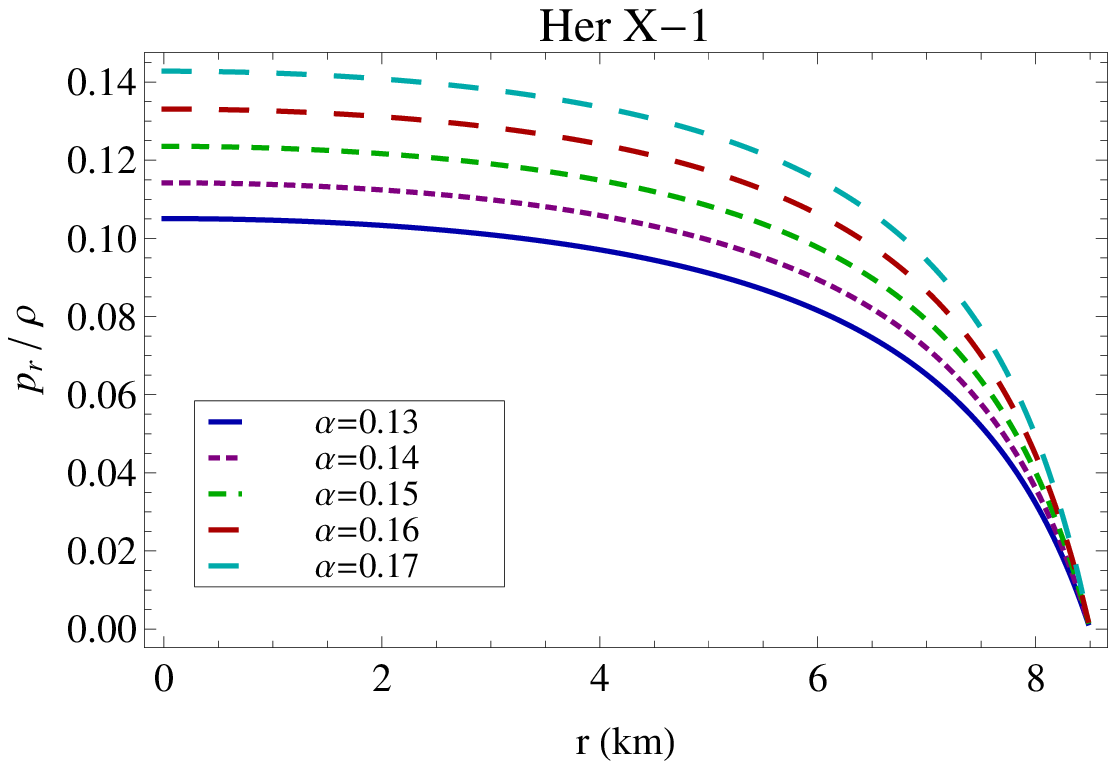}
         \includegraphics[scale=.5]{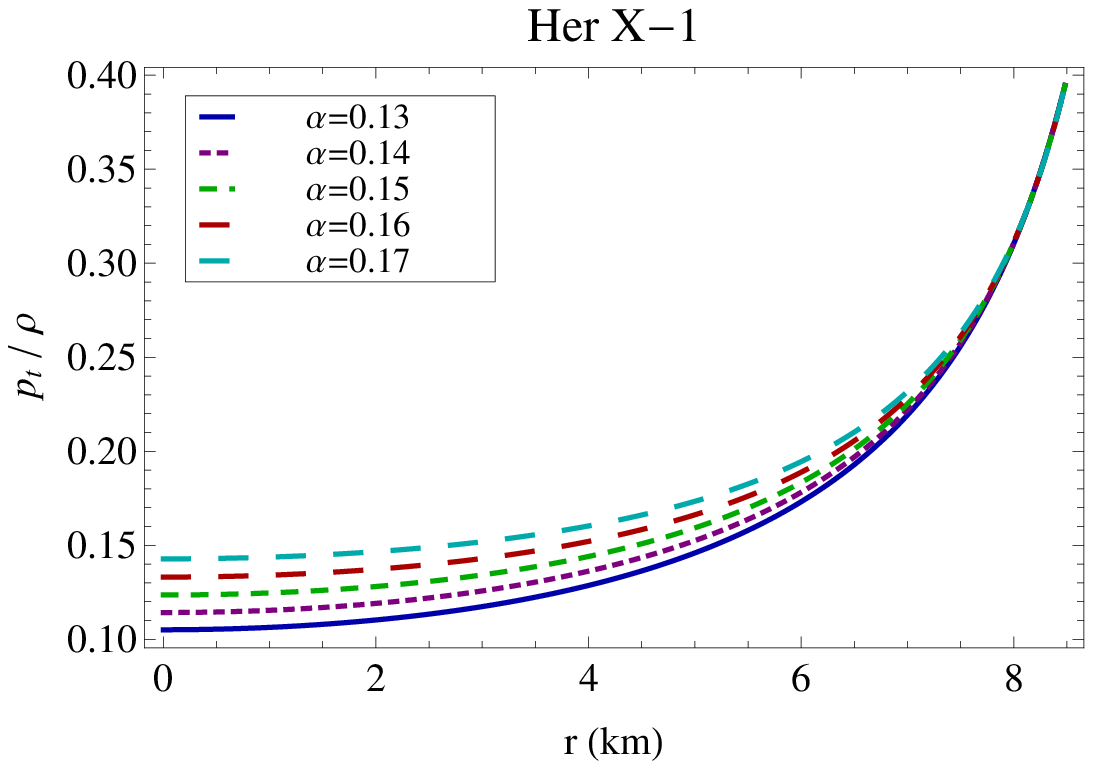}
       \caption{The variation of $p_r/\rho$ and $p_t/\rho$ are shown against radius $r$ for different values of $\alpha$ mentioned in the figure.\label{eos1}}
\end{figure}

\subsection{Causality condition and Cracking method}
In this subsection, we are interested to calculate the sound velocity, which can reflect
the stiffness of the system. According to the definition, the
radial and transverse sound velocities of the system are calculated as,
\begin{eqnarray*}
V_r=\sqrt{\frac{dp_r}{d\rho}}=\sqrt{\frac{p_r'}{\rho'}},\,
V_t=\sqrt{\frac{dp_t}{d\rho}}=\sqrt{\frac{p_t'}{\rho'}},
\end{eqnarray*}
here $V_r$ and $V_t$ respectively denote the radial and transverse speed of sound. We calculate the square of radial and transverse velocity of sound from our present model as,
\begin{eqnarray}
V_r^2&=&\frac{dp_r}{d\rho} =\alpha,\\
 V_t^2&=& \frac{dp_t}{d\rho} = \frac{C_8 + C_9 r^2 +
  3 b C_{10} r^4 +
   b \big\{3 a (1 + \alpha) b - 4 b B + a (1 - 3 \alpha) B^2\big\} r^6 +
  b^2 C_7 r^8}{2 \big[C_3 + C_4 r^2 +
    3 b C_5 r^4 + 6 b^2 (a - B) r^6 + 2 b^3 r^8\big]},
\end{eqnarray}
where $C_8,\,C_9$ and $C_{10}$ are constants given as,
\begin{eqnarray*}
C_8&=&-a^2 + 11 a^2 \alpha + b - 11 \alpha b + a B - 9 a \alpha B - B^2 +
  3 \alpha B^2,\\
  C_9&=&a^3 (1 + \alpha) + a (-5 + 31 \alpha) b +
     4 (1 - 6 \alpha) b B + a (-1 + 3 \alpha) B^2 -
     a^2 (B + 3 \alpha B),\\
     C_{10}&=&a^2 (1 + \alpha) + 2 (-1 + 5 \alpha) b - a (B + 3 \alpha B).
\end{eqnarray*}
\begin{figure}[htbp]
    %\centering
        \includegraphics[scale=.5]{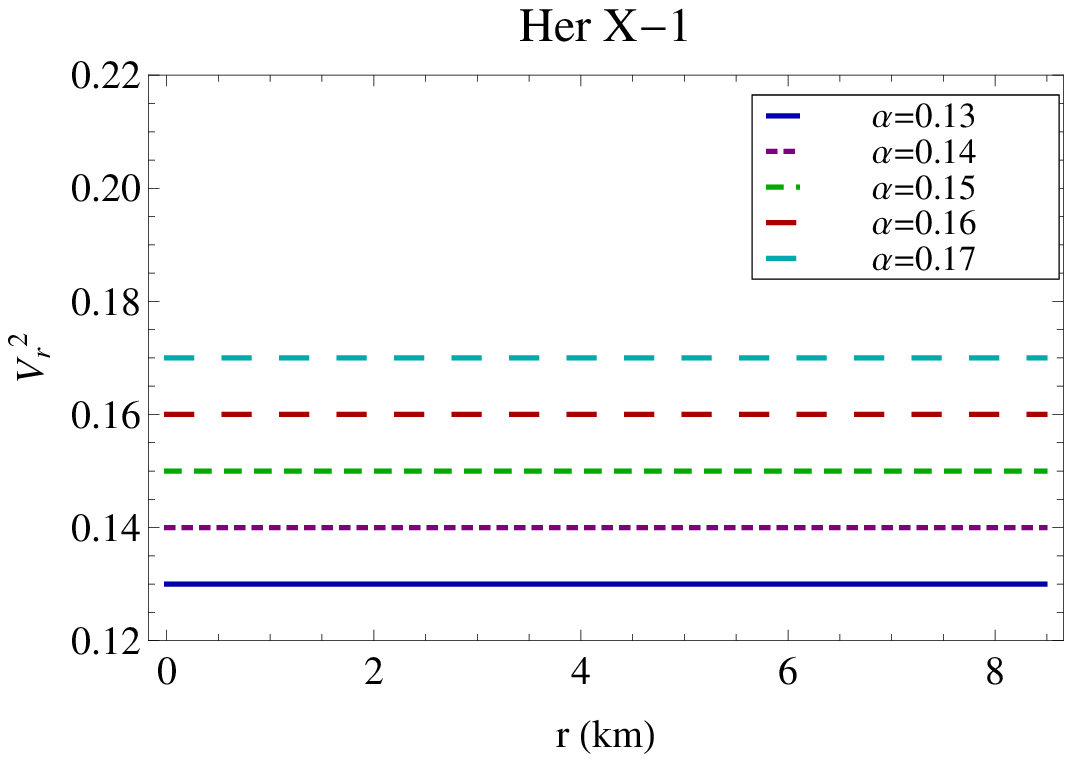}
         \includegraphics[scale=.5]{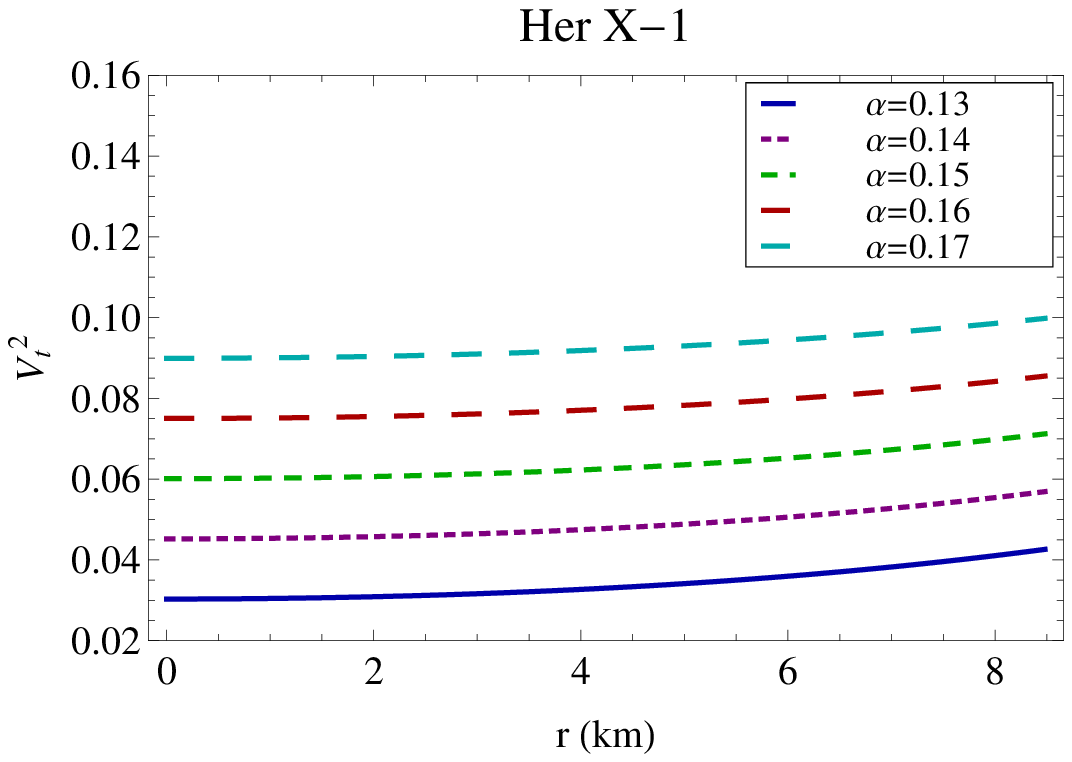}
          \includegraphics[scale=.5]{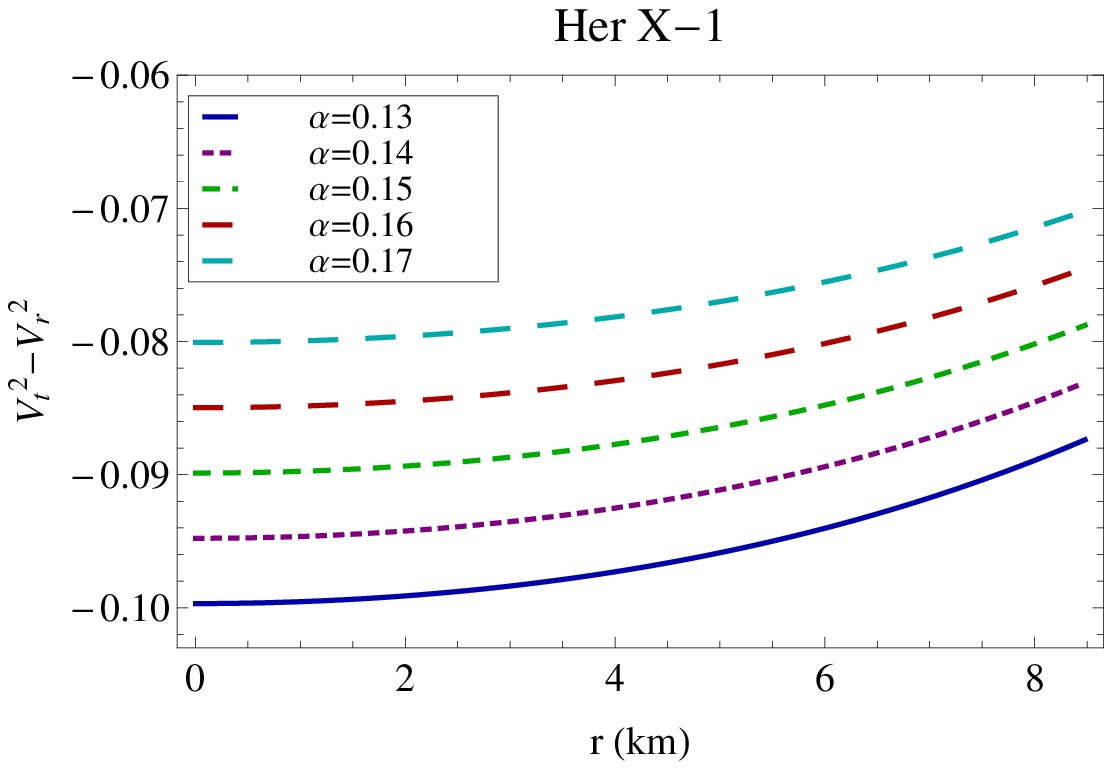}
           \includegraphics[scale=.5]{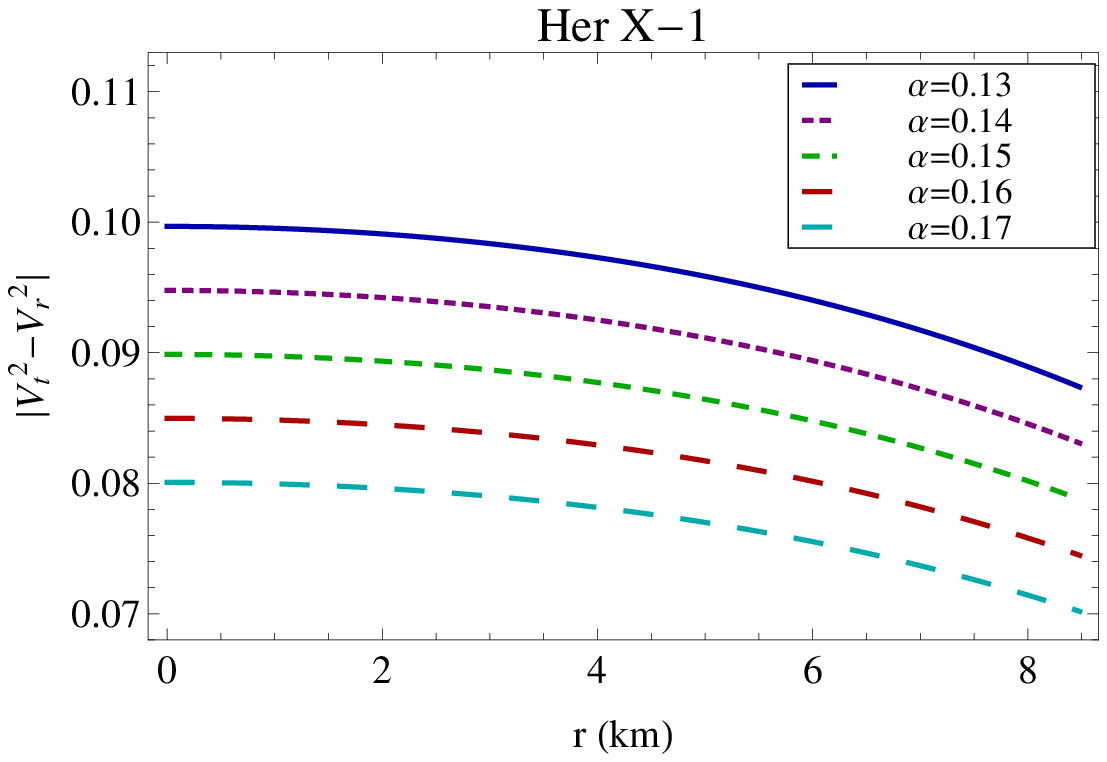}
       \caption{ Square of the radial sound velocity $V_r^2$,  square of the transverse sound velocity $V_t^2$, $V_t^2-V_r^2$ and  the stability factor $|V_t^2-V_r^2|$ are plotted against $r$ inside the stellar interior for Her X-1 for different values of $\alpha$ mentioned in the figures.\label{sv1}}
\end{figure}
The profiles of $V_r^2$ and $V_t^2$ for different values of $\alpha$ are shown in Fig.~\ref{sv1}. The sound velocity is actually concerned with the slope of the $p_r(\rho)$ and $p_t(\rho)$
functions. In principle, the sound speed should be smaller
than the speed of light, and a smaller sound velocity corresponds to a
softer equation of state (EoS) \cite{Li:2017xlb}. From the graphical analysis, we observe that $0<V_r^2,\,V_t^2<1$. Moreover we see that our model of hybrid star is potentially stable since the radial velocity of sound is always dominating the transverse velocity of sound (from Fig.~\ref{sv1}, $V_t^2-V_r^2<0$) in everywhere inside the stellar interior \cite{herrera1992cracking}, also $|V_r^2-V_t^2|<1$ for $0\leq r \leq r_b$ which verifies that there is no cracking in the interior of the star \cite{andreasson2009sharp}.
\subsection{TOV equation}
\begin{figure}[htbp]
    %\centering
        \includegraphics[scale=.4]{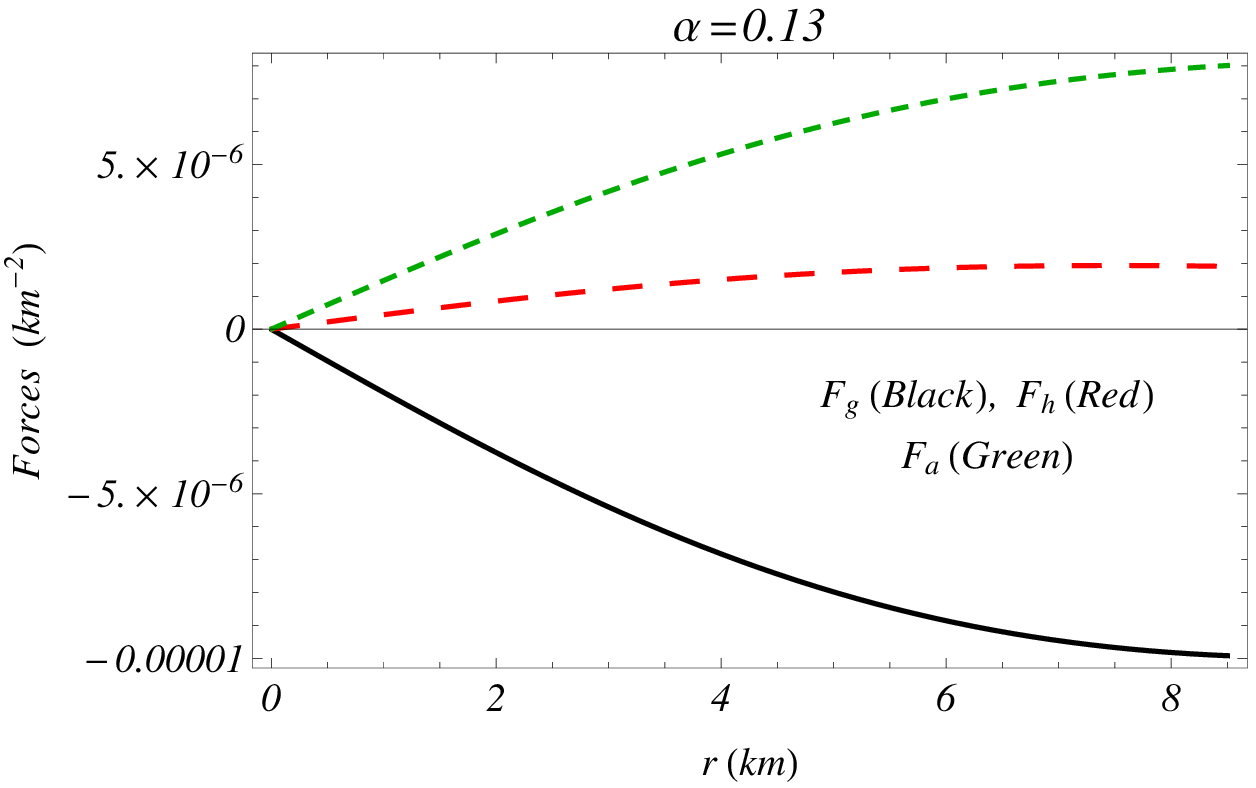}
         \includegraphics[scale=.4]{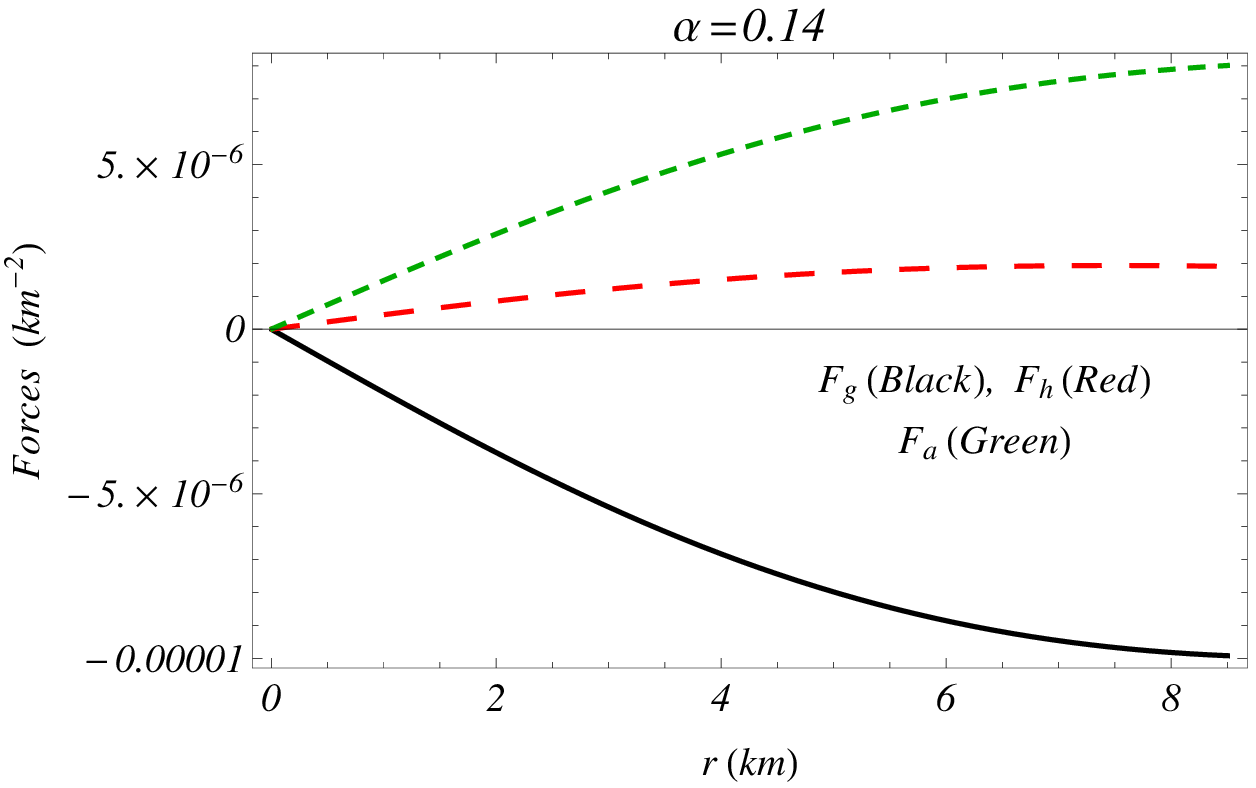}
          \includegraphics[scale=.4]{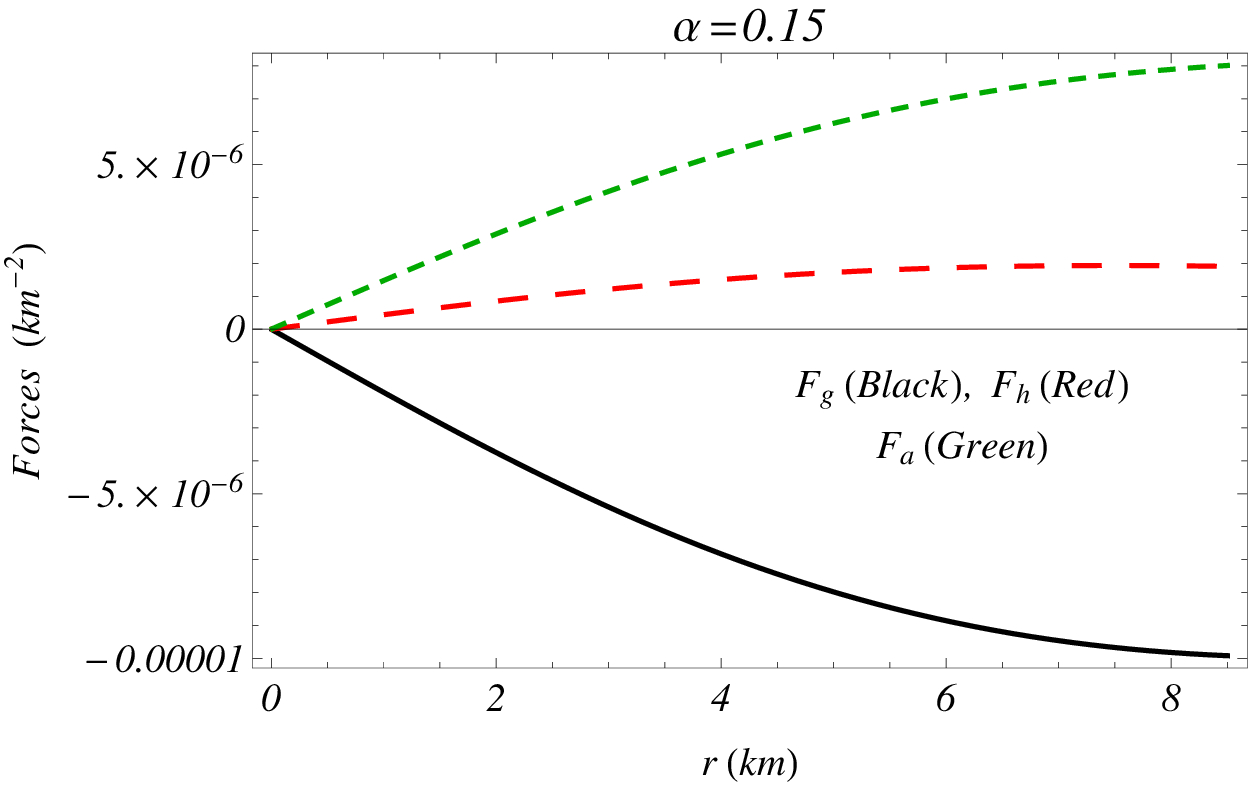}
           \includegraphics[scale=.4]{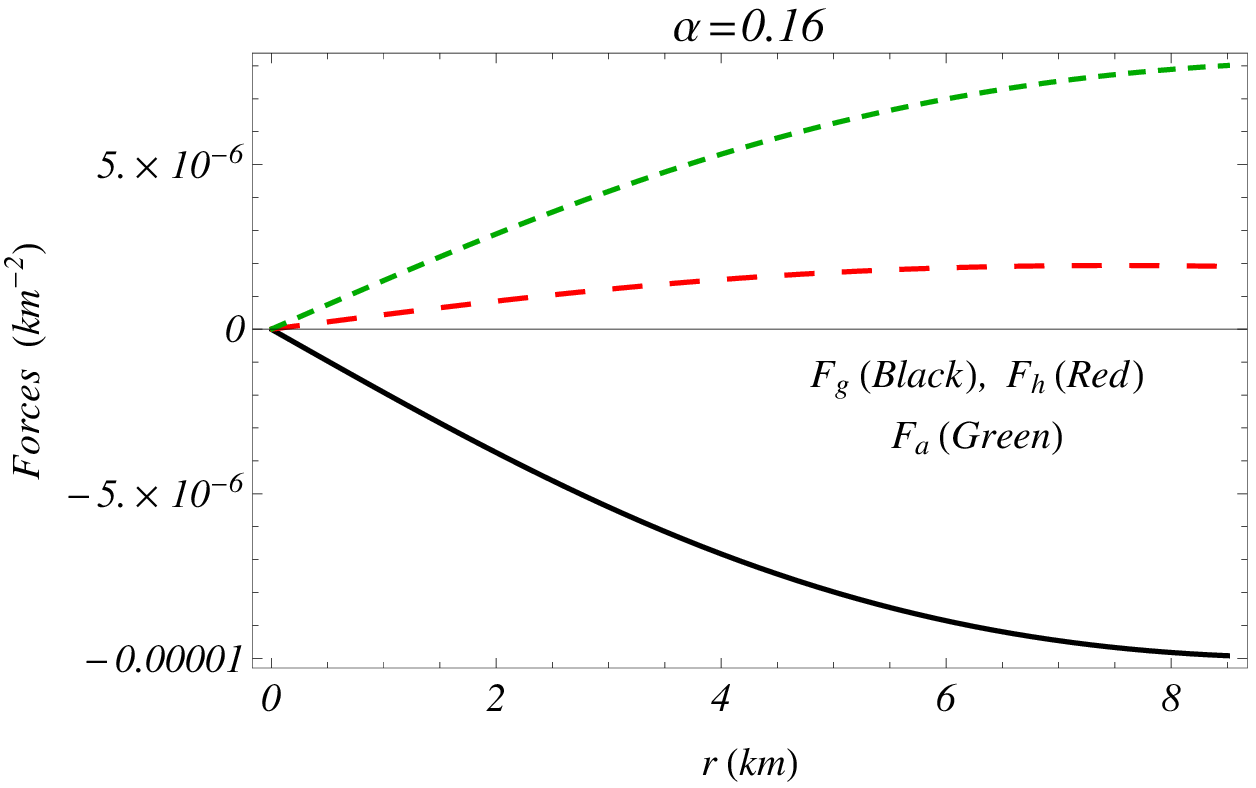}
            \includegraphics[scale=.4]{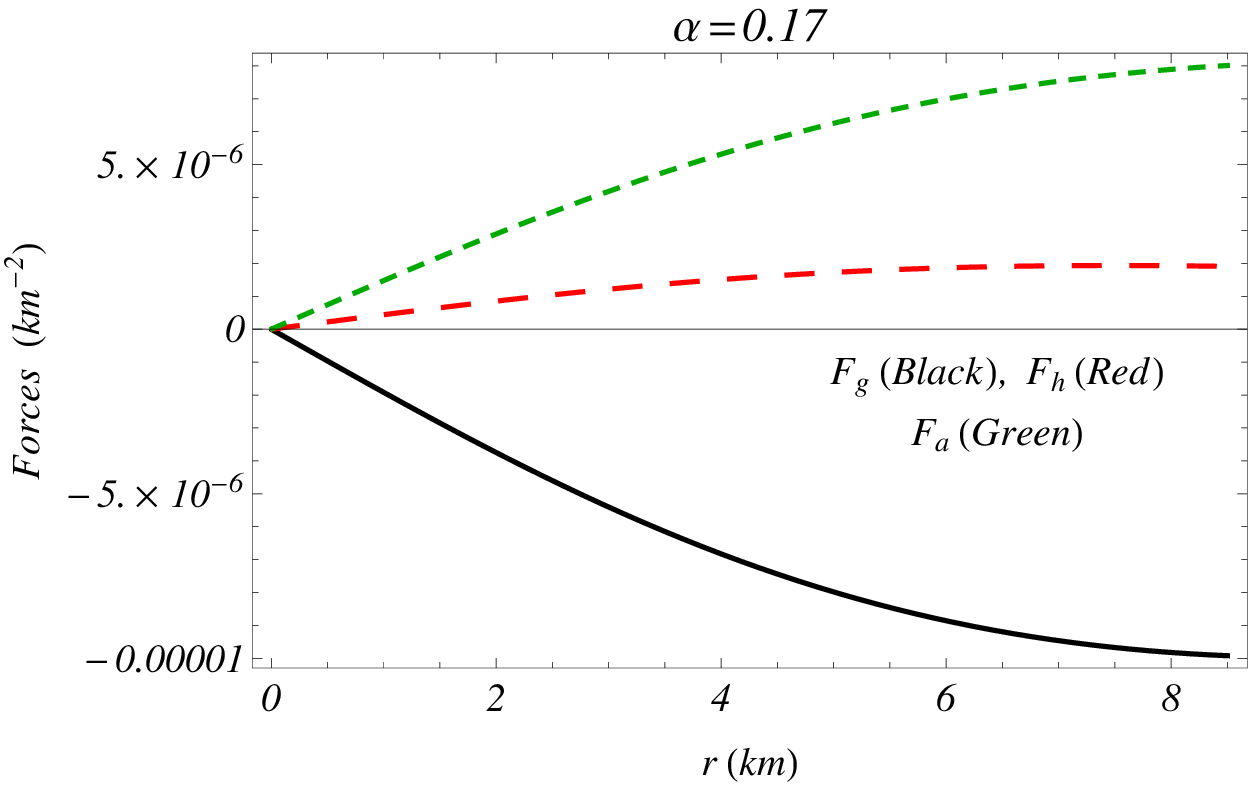}
       \caption{Different forces acting on the present model are plotted against $r$ for Her X-1 by taking different values of $\alpha$.\label{force1}}
\end{figure}

We know that the equilibrium of a gravitationally
bounded fluid configuration in absence of dissipative effects such as heat flow is characterized by the effect of the gravitational force, $F_g$, the hydrostatic force, $F_h$ and the force due to anisotropy, $F_a$. The generalized TOV equation for our present model is described by,
 \begin{eqnarray}
  -\frac{1}{2}\nu'(\rho^{\text{eff}}+p_r^{\text{eff}})-\frac{d}{dr}p_r^{\text{eff}}+\frac{2}{r}(p_t^{\text{eff}}-p_r^{\text{eff}})=0.
\end{eqnarray}
The above equation can be expressed as,
\begin{eqnarray}
F_g+F_h+F_a=0,
\end{eqnarray}
where,
\[F_g=  -\frac{1}{2}\nu'(\rho^{\text{eff}}+p_r^{\text{eff}}),\,F_h=-\frac{d}{dr}p_r^{\text{eff}},\,F_a=\frac{2}{r}(p_t^{\text{eff}}-p_r^{\text{eff}}).\]
The three different forces acting on the stellar system are shown in Fig.~\ref{force1} for different values of $\alpha$. From the figures one can note that the gravitational force is dominating is nature which balances the combine effect of hydrostatics and anisotropic forces to keep the system in equilibrium.
\subsection{Relativistic adiabatic index and Harrison-
Zeldovich-Novikov condition}
\begin{figure}[htbp]
    %\centering
        \includegraphics[scale=.5]{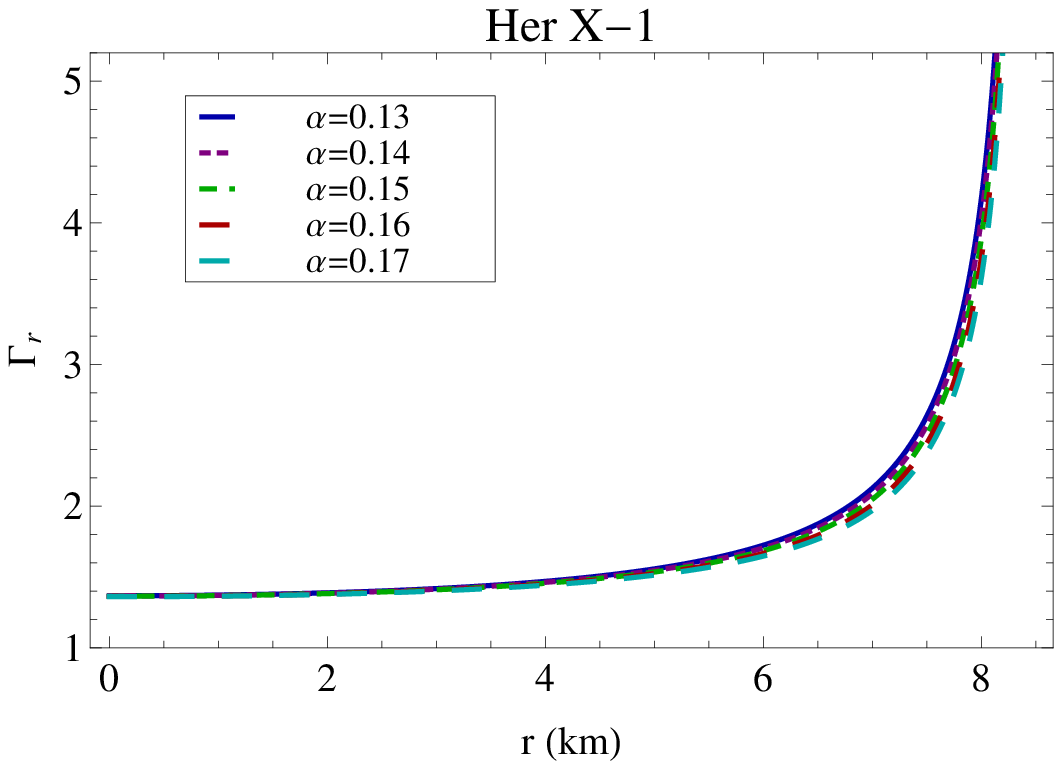}
         \includegraphics[scale=.5]{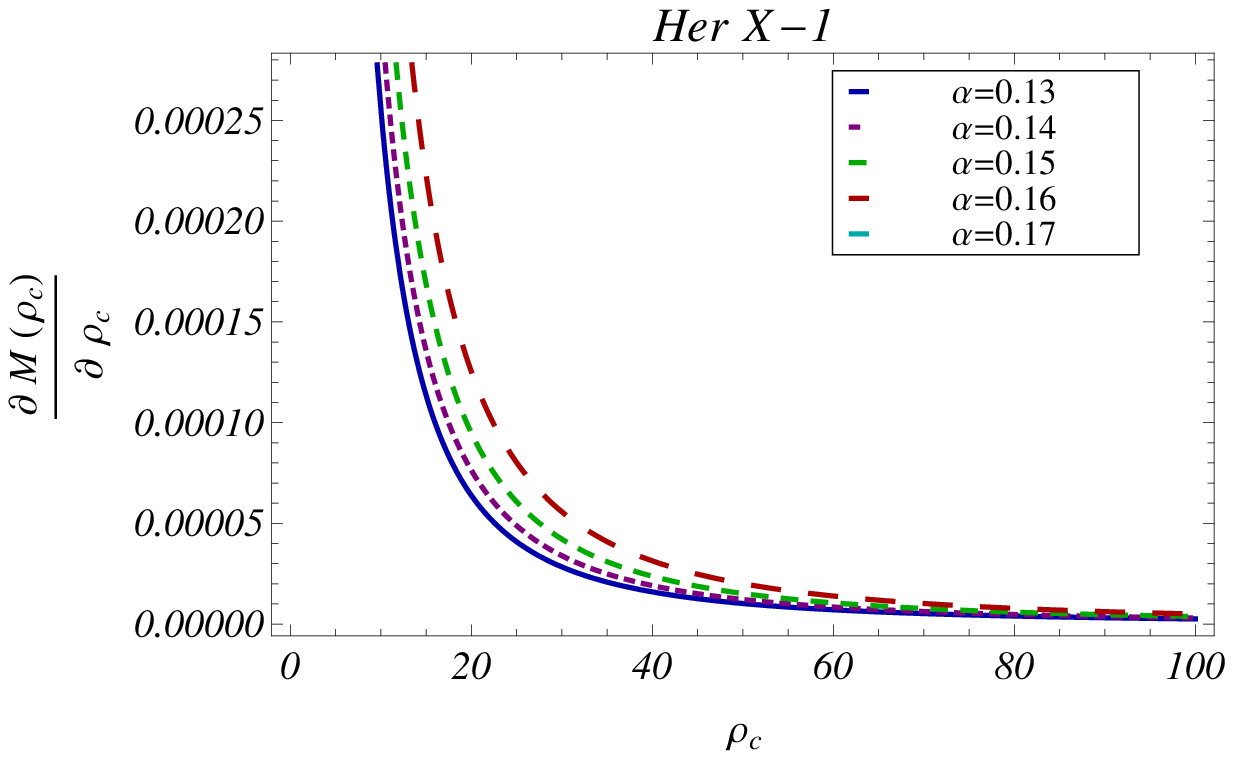}
       \caption{The relativistic adiabatic index and $\frac{\partial M}{\partial \rho_c}$ are shown for different values of $\alpha$ mentioned in the figure.\label{stab1}}
\end{figure}

The adiabatic index, first proposed by Chandrasekhar \cite{Chandrasekhar:1964zz}, determines the stability of a spherical object. Chan et al. \cite{Chan1993} proposed the definition of adiabatic index for isotropic fluid as, $\Gamma=\frac{\rho+p}{p}\frac{dp}{d\rho}$ and for anisotropic fluid sphere this expression changes as, $\Gamma_r=\frac{\rho+p_r}{p_r}\frac{dp_r}{d\rho}$. $\Gamma>\frac{4}{3}$ is said to give the condition of stability for a Newtonian fluid sphere \cite{bondi1964contraction}. We shall use graphical analysis to check this condition because of the expression's intricacy in our present paper. The profile of $\Gamma_r$ is plotted in Fig.~\ref{stab1} for our present model and we see that $\Gamma_r$ is monotonic increasing function of `r'. The numerical values of $\Gamma_{r0}$ for different values of $\alpha$ is presented in the Table~\ref{table4} and it can be seen that $\Gamma_{r0}>4/3$. Due to the monotonic increasing nature of this curve $\Gamma_r>4/3$ everywhere inside the stellar interior.

\begin{table}[t]
\centering
\caption{The numerical values of $\Gamma_{r0}$ for different values of $\alpha$ for the compact star Her X-1 by assuming mass $= 0.85~M_{\odot}$, radius $= 8.5 $ km., $b=0.4\times 10^{-5}$~km$^{-4}$, $B_g=60 MeV/fm^{3}$.}\label{table4}
\begin{tabular}{@{}ccccccccccccc@{}}
\hline
$\alpha$&& $\Gamma_{r0}$  \\
\hline
0.13&& 1.36731  \\
0.14&& 1.36564   \\
0.15&& 1.36397    \\
0.16&& 1.36230       \\
0.17&& 1.36063\\
\hline
\end{tabular}
\end{table}
Next we are interested to check the stability of the present model under Harrison-
Zeldovich-Novikov condition. A stellar model will be unstable if $\frac{\partial M}{\partial \rho_c}<0$ \cite{Zeldovich,Harrison1965}, To check this condition the profile of $\frac{\partial M}{\partial \rho_c}$ has been depicted in Fig.~\ref{stab1}. The graphical behavior shows that $M(\rho_c)$ is an increasing function of $\rho_c$, i.e., $\frac{\partial M}{\partial \rho_c}>0$ everywhere inside the stellar interior and hence it ensures that our proposed model of hybrid star is physically realistic.
\section{Discussion}\label{discussion}

%The quark phase is modeled according to the MIT bag model proposed by Chodos et al. \cite{chodos}. The current masses of up and down quarks are extremely small and are assumed to be 5 and 10 MeV, respectively. For the bag model the energy density and the pressure are given by  \par

%\bibitem{chodos}A. Chodos, R. L. Jaffe, K. Johnson, C. B. Thorn, and V. F. Weisskopf, Phys. Rev. D 9, 3471 (1974).

This paper elaborately describes the influence of MIT bag constant on the
different model parameters of anisotropic hybrid star candidates in Einstein's gravity. The model has been developed on the choice of the metric functions proposed by
Tolman and Kuchowicz \cite{Tolman1939,Kuchowicz1968} and the metric potentials contain the arbitrary constants $a,\,b,\,B$ and $D$. Using the observed
values of masses and radius of six different compact stars mentioned in Table~\ref{table1}, we have successfully obtained the values of the above mentioned arbitrary constants from the boundary conditions. For graphical analysis, we have considered the compact star Her X-1 which was detected by the Uhuru satellite in 1972 \cite{tananbaum1974compact} and identified
with the variable star HZ Her \cite{kurochkin1972peremennye,cherepashchuk1972efremov}. Abubekerov et al. \cite{Abubekerov:2012yj} obtained the first estimates of the masses of the components of the Her X-1/HZ
Her X-ray binary system by taking into account non-LTE effects in the formation of the $H_\gamma$
absorption line and the estimates of mass were made in a Roche
model based on the observed radial-velocity curve of the optical star, HZ Her.\par
From our analysis we have shown that the parameter $\beta$ present in the EoS of normal baryonic matter depends on $\alpha$ and the bag constant $B_g$. The numerical values of $\beta$ for different values of $\alpha$ for Her X-1 have been obtained for the bag constants $B_g=60~MeV/fm^3$ and $B_g=70~MeV/fm^3$ in Tables \ref{table2} and \ref{table3} respectively. From these two tables one can note that the numerical values of $\beta$ decreases when $B_g$ increases for a fixed compact star. It can also be noted that for a particular compact star, when $B_g$ is fixed, the value of $\beta$ increases with the increasing value of $\alpha$. The numerical values of central density and central pressure increase with the increasing value of $\alpha$ for a fixed value of $B_g$. If the value of $B_g$ increases then the numerical value of central density takes lower value corresponding to the same value of $\alpha$. An interesting feature to be noted that the numerical values of surface density do not depend on $\alpha$. On the other hand, the changes of the bag constant $B_g$ do not affect on the central pressure. In both cases (Table~\ref{table2} and Table~\ref{table3}) the central pressure $p_c$ takes equal value for same $\alpha$ when the bag constant $B_g$ varies. It can also  be noticed that, the numerical values of both compactness factor and surface resdshift increase with the increasing value of $\alpha$. The numerical values of $\mathcal{U}$ lie in the range $(0,\,\frac{4}{9})$, the prediction proposed by Buchdahl \cite{Buchdahl1959}. In the absence of a cosmological constant the surface redshift ($z_s$) lies in the range
$z_s \leq 2$ as found in Refs. \cite{Buchdahl1959,straumann2012general,bohmer2006bounds}. For our present model hybrid star $z_s<0.08$ for $B_g$=60 MeV/fm$^{3}$ and $z_s<0.06$ for $B_g$=70 MeV/fm$^{3}$ for different values of $\alpha$ mentioned in the Table \ref{table2} and \ref{table3}. Also, $\Delta=p_t-p_r$ is shown in Fig.~\ref{pr5} for our current model. As can be seen from the figure, there is no variation in the anisotropic factor for different values of $\alpha$, and all profiles are identical. The profile of quark matter density is dominated with increasing value of $\alpha$ and positive throughout the interior of the star, whereas the pressure related to quark matter assumes negative values. The profiles of density and pressure due to normal baryonic matter shows the usual behavior. The equation of state parameters $\omega_r$ and $\omega_t$ both lie in $(0,\,1)$ which indicates radiating era \cite{Sharif:2016xbn}. From the graphical interpretation of mass-radius relation of compact star
we observe that with the larger values of $\alpha$, the stellar system turns into a more massive compact object (Fig.~\ref{mr1}). One can observe that
all energy bounds are fulfilled for considered stellar model which guarantees
the existence of ordinary matter inside the stellar interior. We have also obtained the maximum allowable mass for different values of $\alpha$ from our present analysis in Fig.~\ref{mr2} and discovered that as the value of $\alpha$ increases, the star becomes more massive. It's also worth noting that the square of both the radial and transverse sound speeds is less than $1$, moreover the potentially stable condition is satisfied since $V_t^2<V_r^2$ everywhere inside the stellar model. The forces associated with the TOV equation in Fig.~\ref{force1} indicate the balancing nature of the system for different values of $\alpha$. Other model stability conditions have been studied analytically and are depicted in Fig.~\ref{stab1}. \par
So, in conclusion, this present model describes some important features of hybrid star in (3+1)-dimensional analysis and in the context of general relativity, one may check that the compact star meets all of the physically acceptable parameters in the range
$0.13\leq \alpha \leq 0.17$ for $B_g =$ 60 MeV/fm$^3$. {\bf It was demonstrated that the model satisfies the limitations imposed by gravitational data.} Also
the behavior of the stars can be checked for another range of $\alpha$ for the suitable choice of the model parameter and the bag constant $B_g$. {\bf In the future, it will be fascinating to see if EoS for hybrid stars can explain neutron star cooling, magnetic fields, and faults in hybrid stars.}Therefore, within the framework of general relativity, this model could be utilized to describe hybrid stars.

\section*{Acknowledgements} P.B. is thankful to the Inter University Centre for Astronomy and Astrophysics (IUCAA), Pune, Government of India, for providing visiting associateship. PB is also thankful to Department of Science \& Technology and Biotechnology, Government of West Bengal  for providing research grant (Memo No: STBT-11012(26)/23/2019-ST SEC).

\bibliographystyle{unsrt}
\bibliography{hybrid_final}

\end{document}